\documentclass[showpacs, pra,twocolumn,preprintnumbers ,amsmath, amssymb, superscriptaddress, aps]{revtex4-2}
\usepackage{color}
\usepackage{amsmath,amssymb}
\usepackage{pifont}
\usepackage{amssymb}  
\usepackage{bbold}
\usepackage{float}
\usepackage{placeins}
\usepackage{subfloat}
\usepackage{adjustbox}

\usepackage[caption=false]{subfig}
\usepackage{tikz}
\usepackage{makecell}
\usepackage{subfig}
\usepackage{pifont}  
\usepackage{graphicx}

\graphicspath{{Figures/}}
\usepackage{dcolumn}  
\usepackage{bm}       
\usepackage{multirow} 
\usepackage[colorlinks]{hyperref}
\usepackage{mathtools}
\usepackage{appendix}

\def \be{\begin{align}}
	\def \ee{\end{align}}
\def \bea{\begin{eqnarray}}
	\def \eea{\end{eqnarray}}

\begin{document}

	\title{Magnetic control of electron scattering in silicene quantum dots}

	\date{\today}
	
	\author{Mohamed El Azar}
	\email{elazar.m@ucd.ac.ma}
	\affiliation{Laboratory of Theoretical Physics, Faculty of Sciences, Choua\"ib Doukkali University, PO Box 20, 24000 El Jadida, Morocco}
	
	\author{Elmustapha Feddi}
	
	\affiliation{ School of Applied and Engineering Physics, University Mohammed VI Polytechnic, Ben Guerir, 43150, Morocco }
	
	\author{Pablo Díaz}
	\affiliation{Departamento de Ciencias F\'{i}sicas, Universidad de La Frontera, Casilla 54-D, Temuco 4811230, Chile}  
	\author{David Laroze}
	\affiliation{Instituto de Alta Investigación, Universidad de Tarapacá, Casilla 7D, Arica, Chile}
	\author{Ahmed Jellal}
	\email{a.jellal@ucd.ac.ma}
	\affiliation{Laboratory of Theoretical Physics, Faculty of Sciences, Choua\"ib Doukkali University, PO Box 20, 24000 El Jadida, Morocco}


	\begin{abstract}
		The Klein tunnel effect phenomenon makes it impossible to permanently confine charge carriers in massless nanostructures. However, applying a constant magnetic field allows these electrons to be temporarily localized, thus forming quasi-bound states. In this study, we analyze the mechanism of electron diffusion through a silicene quantum dot (SQD) subjected to a perpendicular magnetic field. To enhance spatial localization, we exploit the spin-orbit coupling (SOC) specific to silicene, which generates a natural energy gap by acting as an effective mass. We first derive the solutions to the Dirac equation at low energy. Subsequently, by imposing the continuity conditions at the SQD interfaces,  we obtain exact expressions for the diffusion coefficients. These results are then used to map the scattering efficiency together with the spatial distributions of probability and current densities. Our simulations demonstrate that the presence of this intrinsic gap significantly enhances electron trapping at the center of the SQD. Finally, we prove that the interplay between the external field and SOC breaks spin symmetry, thereby enabling robust and spin-selective confinement.
	\end{abstract}
	
	\maketitle

	\section{Introduction}

	Charge carriers confined within nanometric dimensions 
	experience a profound alertation of their energy landscape. The continuous energy bands characteristic of bulk materials collapse into discrete, "artificial atom"  levels due to the compression of electron wavefunctions.  
	The basic form of quantum confinement enables one to control three specific properties, including optical selection rules, spin degeneracy, and energy spacing. The quantization technology developed today serves as the foundation for nanoelectronics and new quantum systems, allowing progress in three areas: single-electron measurement systems, and spin-based solid-state qubits, and high-fidelity photon-matter interfaces.\cite{Loss1998,Hanson2007,Zwanenburg2013,Kloeffel2013,Qammar2024}. Quantum dots thus emerge as highly controlled platforms. Operating at a scale where boundary conditions, fundamental symmetries, and topological characteristics manifest as measurable signatures, bridging the behavior of mesoscopic devices with the principles of microscopic quantum mechanics \cite{Hanson2007,Zwanenburg2013}.
	Condensed matter physics gains previously unheard-of structural richness from two-dimensional Dirac materials. Their low-energy charge carriers, governed by an effective Dirac equation, exhibit a linear dispersion relation near band contact points.  Remarkable electronic optical phenomena, including collimation, Veselago-type refraction, and chirality-protected transport, are made possible by this special topology and are entirely unattainable in conventional semiconductors with parabolic dispersion \cite{CastroNeto2009,Katsnelson2006}. 
	Graphene exemplifies this class of materials because its quasiparticles behave as massless fermions whose pseudospin is rigidly coupled to momentum.
	This physical characteristic turns out to be a double-edged sword: although it provides outstanding ballistic transport qualities, it significantly impedes traditional confinement techniques \cite{CastroNeto2009,Katsnelson2006}.
	The challenge of confinement in graphene is well documented. In pure graphene, the chiral nature of Dirac fermions imposes perfect transmission under normal incidence through electrostatic barriers—a famous phenomenon known as the Klein tunneling effect. Consequently, quantum dots defined solely by electrostatic grids remain intrinsically porous, regardless of the height of the barrier applied \cite{Katsnelson2006,Stander2009}. Experiments conducted on grid-controlled junctions have formally validated this Klein-type transport and its strong angular dependence \cite{Stander2009}. Nevertheless, recent studies have nuanced this understanding. In junction engineering, it has been shown that judicious exploitation of the pseudo-spin and the geometry of the device can lead to a marked suppression of transmission (anti-Klein behavior). These advances highlight the complexity and subtlety of the interaction between chirality, mode matching, and interface effects \cite{Banszerus2020,Elahi2024}.

	In order to overcome the persistent challenge posed by the Klein tunneling effect, several confinement architectures have been developed. These aim to artificially break chiral symmetry or geometrically modify electron trajectories relative to  normal incidence. Although physical structuring, through edge termination or infinite potential walls, can effectively localize electronic states, this approach has intrinsic limitations in terms of disorder sensitivity. In addition, it induces detrimental intervalley scattering \cite{Downing2011,Romanovsky2012}. 
	Electrostatic quantum dots in graphene can also host long-lived quasi-bound states, commonly referred to as “resonances.” This temporary trapping occurs in two scenarios: either when the underlying classical dynamics are sufficiently integrable, or when the potential profile creates effective barriers for specific kinetic momenta \cite{Matulis2008,Bardarson2009}. However, a more robust approach involves introducing a local mass term. The opening of an energy gap in the Dirac spectrum results in the lifting of chirality-protected transmission, thereby allowing true electrostatic confinement \cite{Cheianov2006,Elahi2024}.
	These theoretical paradigms have stimulated intense experimental efforts, leading to the fabrication of Dirac quantum dots via scanning tunneling microscope (STM) lithography and tip-induced potentials. Such platforms, characterized by a high degree of structural and electronic purity, enable the exploration of symmetry breaking, spectral tunability, and robust particle confinement with atomic-scale precision \cite{Gutierrez2016}.
	
	The application of an external magnetic field acts as a particularly powerful and direct control parameter on these charge carriers. By forcing semi-classical trajectories to adopt cyclotron orbits, the magnetic flux induces a fundamental restructuring of the energy spectrum through Landau quantization. In pure graphene, magnetic confinement manifests itself as an anomalous quantum Hall effect, which constitutes a direct macroscopic signature of the underlying Dirac dispersion and the non-trivial Berry phase of Dirac fermions \cite{Novoselov2005,Zhang2005}. The extreme sensitivity of Dirac electrons to magnetic flux becomes particularly apparent when the material is placed on a periodic potential or a moiré superlattice. The complex competition between the magnetic length and the lattice period leads to the emergence of fractal energy spectra, characteristic of Hofstadter butterfly physics \cite{Dean2013}. At the mesoscopic scale, this sensitivity to magnetic flux governs transport within confined geometries, and gives rise to the Aharonov-Bohm interference patterns, known for their high robustness. This demonstrates that quantum phase coherence and topology become directly accessible experimental observables \cite{Schelter2012,Romanovsky2012}.

	Silicene consists of 2D silicon allotropes and displays a group-IV element similarity to graphene. It exhibits a slightly buckled honeycomb lattice rather than a perfectly planar structure. This intrinsic structural distortion provides a highly versatile platform for extending confinement strategies \cite{Cahangirov2009}. Analysis of this buckled lattice configuration reveals an enhanced intrinsic spin–orbit coupling (SOC) while preserving the two-valley Dirac electronic structure.
	However, the introduction of a perpendicular electric field induces sublattice symmetry breaking, generating a tunable energy gap whose amplitude can be precisely controlled \cite{Liuspin2011, Drummond2012}. The resulting effective mass term is of crucial physical importance, as it suppresses the Klein tunneling effect and enables robust spin- and valley-selective localization channels. By exploiting this property, contemporary models have shown that gate-defined mass profiles in silicene quantum dots (SQDs) can effectively trap charge carriers \cite{Szafran2018}.
	In addition, this well-controlled confinement regime allows the modulation of infrared optical responses and opens pathways toward advanced spin-dependent transport architectures in silicene nanoribbon devices \cite{Chen2024EPL, Gao2024APL}.
	
	In the context of the urgent need for robust confinement architectures, our study builds upon previous works, notably  \cite{Pena2022, Elazarmasse}, to theoretically investigate the complex interplay between an external magnetic field and intrinsic spin–orbit coupling (SOC) in silicene. We focus on electron scattering through a circular silicene quantum dot (SQD), a system where the combination of lattice buckling and SOC naturally introduces a tunable energy gap.
	{While pure magnetic confinement in pristine graphene has been shown to induce remarkable scattering phenomena, such as the formation of caustical wave patterns and quasi-bound states \cite{garg2022caustical}, the well-known Klein tunneling effect  inherently limits carrier confinement in such gapless systems. In contrast, the SOC-induced gap in silicene serves as an effective mass barrier, suppressing electron leakage and enabling highly controlled carrier localization.} 
	This property enables the generation of stable, spin- and valley-polarized confinement channels that can be directly manipulated through external fields.
	By analytically solving the low-energy Dirac equation, we systematically evaluate key physical observables, including scattering efficiency, the spatial distribution of probability densities, and the formation of closed current vortices within the quantum dot. We also explore the emergence of resonant states, their dependence on the dot size, magnetic flux, and SOC strength, and the resulting modifications of the local density of states.
	Our study goes beyond  single-particle confinement, demonstrating how the proposed system enables control of spin- and valley-dependent transport. Our results show that SQDs can function as versatile platforms for quantum devices that exploit  magnetic and spin-orbit coupling (SOC) effects to control charge, spin, and valley degrees of freedom with high precision. In this way, our study bridges basic science and potential applications in spintronics, valleytronics, and quantum information technologies.

	The present paper is organized as follows. In Sec. \ref{sec:theory}, we provide an in-depth analysis of the underlying theoretical framework. In fact, we establish the theoretical foundation of our system and details the analytical derivation of Dirac spinors. In Sec. \ref{scattform}, we focus on the formalism of scattering by establish a rigorous definition of the fundamental physical metrics used to characterize the confinement process. These metrics include scattering efficiency, probability density, and current density. Sec. \ref{sec:results} is devoted to  numerical analysis of the scattering metrics. We specifically highlight how the synergistic effects of the magnetic field and SOC not only significantly extend the lifetimes of resonances, but also lift spin degeneracy, thereby endowing the SQD with intrinsic spatial filtering capabilities. Sec. \ref{sec:conclusions} aims to present a summary of the major conclusions we have reached.

	\section{Theoretical framework}
	\label{sec:theory}

\begin{figure}[ht]
		\centering
		\includegraphics[scale=0.155]{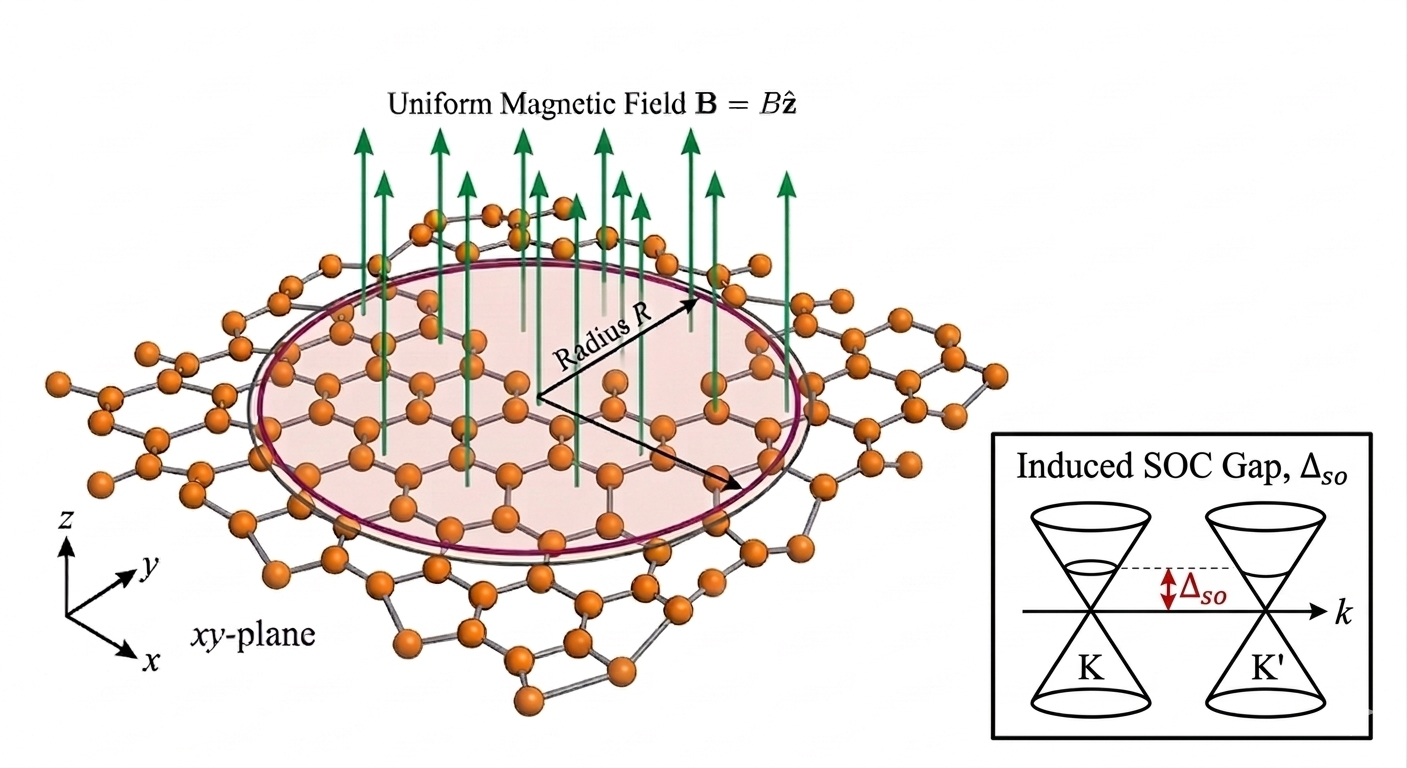}    \caption{A silicene quantum dot of radius $R$ is placed in the horizontal plane $xy$. The magnetic field $B$ is oriented perpendicular to the dot $z$-plane .}
		\label{figsys}
	\end{figure}

	We consider a circular SQD of radius $R$ situated in the $xy$-plane, with a uniform magnetic field $\mathbf{B} = B\hat{z}$ applied perpendicular to the lattice plane, as depicted  in Fig.~\ref{figsys}. The low-energy electronic properties of silicene in the vicinity of the Dirac points $K$ and $K'$ are governed by an effective Hamiltonian that accounts for the buckled lattice structure and the resultant spin-orbit interaction \cite{LiuHall2011,Liu2011}.

	In the absence of external fields, the effective low-energy Hamiltonian for silicene near the $\eta K$ valley ($\eta = +1$ for $K$, $\eta = -1$ for $K'$) takes the form
	\begin{equation}
		\mathcal{H}_0 = \hbar v_F \left( \eta \sigma_x p_x + \sigma_y p_y \right) - \eta s_z \lambda_{SO} \sigma_z,
		\label{eq:H0_silicene}
	\end{equation}
	where $v_F \approx 5.5 \times 10^5$ m/s denotes the Fermi velocity in silicene \cite{CahangirovPRL2009, Kara2012}, $\bm{\sigma} = (\sigma_x, \sigma_y, \sigma_z)$ represents the Pauli matrix vector acting on the sublattice pseudospin degree of freedom, and $\mathbf{p} = (p_x, p_y)$ is the two-dimensional momentum operator. The parameter $\lambda_{SO}$ quantifies the intrinsic SOC strength, while $s_z = \pm 1$ labels the electron spin projection along the $z$-axis. 
	The spin-orbit term $-\eta s_z \lambda_{SO} \sigma_z$ opens a gap at the Dirac points, with the gap magnitude depending on both the valley index and spin orientation. For a given spin-valley combination $(\eta, s_z)$, the effective mass gap is
	\begin{equation}
		\Delta_{SO} = -\eta s_z \lambda_{SO}.
		\label{eq:mass_gap}
	\end{equation}
	This spin-valley locking constitutes a defining characteristic of silicene's electronic structure and underlies its classification as a quantum spin Hall insulator \cite{LiuHall2011}.
	To incorporate the external magnetic field, we employ the minimal coupling  $\mathbf{p} \rightarrow \bm{\pi} = \mathbf{p} + e\mathbf{A}(\mathbf{r})$, with $e > 0$ is the elementary charge of fermion and $\mathbf{A}(\mathbf{r})$ is the magnetic vector potential, chosen in the symmetric gauge 
	%
	$\mathbf{A}(\mathbf{r}) 
	= \frac{Br}{2}\hat{\varphi},$
	%
	where $(r, \varphi)$ denotes polar coordinates centered on the quantum dot.
	The complete Hamiltonian describing a Dirac fermion in silicene subject to the perpendicular magnetic field becomes
	\begin{equation}
		\mathcal{H}(r,\varphi) = v_F \bm{\sigma} \cdot \bm{\pi} + m_{\text{eff}} v_F^2 \sigma_z,
		\label{eq:H_full}
	\end{equation}
	where we have introduced an effective mass term
	%
	$m_{\text{eff}} = -\frac{\eta s_z \lambda_{SO}}{v_F^2},$
	%
	and restrict our attention to a single valley ($\eta = +1$) without loss of generality, as the opposite valley yields analogous results related by time-reversal symmetry.
	In polar coordinates, the Pauli matrices assume the representation \cite{Matulis2008, Recher2007}
	\begin{equation}
		\sigma_r = \begin{pmatrix} 0 & e^{-i\varphi} \\ e^{i\varphi} & 0 \end{pmatrix}, \quad
		\sigma_\varphi = \begin{pmatrix} 0 & -ie^{-i\varphi} \\ ie^{i\varphi} & 0 \end{pmatrix}.
		\label{eq:pauli_polar}
	\end{equation}
	The kinetic term in the polar coordinates reads
	\begin{equation}
		\bm{\sigma} \cdot \bm{\pi} = -i\hbar \left[ \sigma_r \left( \partial_r + \frac{1}{2r} \right) + \frac{\sigma_\varphi}{r} \left( \partial_\varphi - \frac{ier^2 B}{2\hbar} \right) \right].
		\label{eq:kinetic_polar}
	\end{equation}


	The stationary states of the system satisfy the eigenvalue equation
	\begin{equation}
		\mathcal{H}(r,\varphi) \Psi_k(r,\varphi) = E \Psi_k(r,\varphi),
		\label{eq:eigenvalue}
	\end{equation}
	where $E$ denotes the energy eigenvalue and $\Psi_k$ the corresponding two-component spinor wavefunction.
	A crucial observation that facilitates the solution is noting that the total angular momentum operator  
	%
	$J_z 
	-i\hbar \partial_\varphi + \frac{\hbar}{2}\sigma_z$
	%
	commutes with the Hamiltonian, i.e., $[\mathcal{H}, J_z] = 0$. This conservation law permits the factorization of angular and radial dependencies through the ansatz
	\begin{equation}
		\Psi_k(r,\varphi) = \begin{pmatrix} \psi_A^{(k)}(r) e^{il\varphi} \\ i\psi_B^{(k)}(r) e^{i(l+1)\varphi} \end{pmatrix},
		\label{eq:spinor_ansatz}
	\end{equation}
	where $l \in \mathbb{Z}$ is the orbital angular momentum quantum number. The eigenvalue of $J_z$ corresponding to this state is $\hbar(l + 1/2)$.
	Substituting the ansatz \eqref{eq:spinor_ansatz} into \eqref{eq:eigenvalue} yields a coupled system of first-order ordinary differential equations for the radial components
	%
		\begin{align}
			&\frac{d\psi_A^{(k)}}{dr} + \left( \frac{r}{2\ell_B^2} - \frac{l}{r} \right) \psi_A^{(k)} + \left( \alpha k - \frac{m_{\text{eff}} v_F}{\hbar} \right) \psi_B^{(k)} = 0, \label{eq:radial_a} \\
			&\frac{d\psi_B^{(k)}}{dr} - \left( \frac{r}{2\ell_B^2} - \frac{l+1}{r} \right) \psi_B^{(k)} - \left( \alpha k + \frac{m_{\text{eff}} v_F}{\hbar} \right) \psi_A^{(k)} = 0, \label{eq:radial_b}
		\end{align}
		%
		where we have introduced the magnetic length
		$\ell_B = \sqrt{\frac{\hbar}{eB}}$,
		which sets the characteristic length scale for magnetic confinement.
		Decoupling the system by eliminating $\psi_B^{(k)}$ produces a second-order equation for $\psi_A^{(k)}$
		\begin{equation}
			\frac{d^2 \psi_A^{(k)}}{dr^2} + \frac{1}{r}\frac{d\psi_A^{(k)}}{dr} + \left[ k_{\text{eff}}^2 + \frac{l+1}{\ell_B^2} - \frac{r^2}{4\ell_B^4} - \frac{l^2}{r^2} \right] \psi_A^{(k)} = 0,
			\label{eq:second_order}
		\end{equation}
		and the effective wavevector is defined as
		\begin{equation}
			k_{\text{eff}}^2 = k^2 - \frac{m_{\text{eff}}^2 v_F^2}{\hbar^2}.
			\label{eq:k_eff}
		\end{equation}
		%
		
		To identify physically admissible solutions, we examine the asymptotic behavior of \eqref{eq:second_order} in the limits $r \rightarrow \infty$ and $r \rightarrow 0$. %
		In the asymptotic limit where $r \rightarrow \infty$, the differential equation is primarily governed by the $-r^2/(4\ell_B^4)$ term. This dominant contribution allows us to simplify the expression to the following one
		\begin{equation}
			\frac{d^2 \psi_A^{(k)}}{dr^2} + \frac{1}{r}\frac{d\psi_A^{(k)}}{dr} - \frac{r^2}{4\ell_B^4} \psi_A^{(k)} = 0.
			\label{eq:asymp_large}
		\end{equation}
		The substitution $\rho = r^2/(4\ell_B^2)$ transforms this into the modified Bessel equation of order zero, whose normalizable solution behaves as
		\begin{equation}
			\psi_A^{(k)}(r) \sim \frac{\exp\left( -r^2/4\ell_B^2 \right)}{\sqrt{r/2\ell_B}}. 
			\label{eq:asymp_large_sol}
		\end{equation}
		%
		Conversely, as we approach the center of the quantum dot ($r \rightarrow 0$), the behavior of the equation is strictly dictated by the centrifugal barrier $-l^2/r^2$, leading to the following simplified form
		\begin{equation}
			\frac{d^2 \psi_A^{(k)}}{dr^2} + \frac{1}{r}\frac{d\psi_A^{(k)}}{dr} - \frac{l^2}{r^2} \psi_A^{(k)} = 0.
			\label{eq:asymp_small}
		\end{equation}
		Regularity at the origin requires
		\begin{equation}
			\psi_A^{(k)}(r) \sim r^{|l|}. 
			\label{eq:asymp_small_sol}
		\end{equation}
		Incorporating both asymptotic constraints, we posit the ansatz
		\begin{equation}
			\psi_A^{(k,\pm)}(r) = r^{\pm l} \exp\left( -\frac{r^2}{4\ell_B^2} \right) \chi_A^{(k,\pm)}(r),
			\label{eq:ansatz_chi}
		\end{equation}
		where the superscript $+$ ($-$) applies for $l \geq 0$ ($l < 0$), ensuring regularity. The substitution $\zeta = r^2/(2\ell_B^2)$ transforms the equation for $\chi_A^{(k,\pm)}$ into the confluent hypergeometric (Kummer) form
		\begin{equation}
			\zeta \frac{d^2 \xi_A^{(k,\pm)}}{d\zeta^2} + (b_\pm - \zeta) \frac{d\xi_A^{(k,\pm)}}{d\zeta} + a_\pm \xi_A^{(k,\pm)} = 0,
			\label{eq:kummer}
		\end{equation}
		where $\xi_A^{(k,\pm)}(\zeta) = \sqrt{\zeta}\, \chi_A^{(k,\pm)}(\zeta)$, and the parameters are
		\begin{align}
			& a_+ = -\frac{\ell_B^2 k_{\text{eff}}^2}{2},  \quad b_+ = l + 1, \label{eq:params_plus} \\
			& a_- = -l - \frac{\ell_B^2 k_{\text{eff}}^2}{2}, \quad b_- = 1 - l. \label{eq:params_minus}
		\end{align}
		The regular solutions are the confluent hypergeometric functions of the first kind, ${}_1F_1(a; b; \zeta)$. Assembling the complete solution, we find for $l \geq 0$ and $l < 0$, respectively, 
		%
		\label{eq:psiA_solutions}
		\begin{align}
			\psi_A^{(k,+)} &= r^l e^{-r^2/4\ell_B^2} \, {}_1F_1\left( -\frac{\ell_B^2 k_{\text{eff}}^2}{2}; l+1; \frac{r^2}{2\ell_B^2} \right) \label{eq:psiA_plus} \\
			\psi_A^{(k,-)} &= r^{-l} e^{-r^2/4\ell_B^2} \, {}_1F_1\left( -l - \frac{\ell_B^2 k_{\text{eff}}^2}{2}; 1-l; \frac{r^2}{2\ell_B^2} \right). \label{eq:psiA_minus}
		\end{align}
		The components $\psi_B^{(k,\pm)}(r)$ follow from back-substitution into  \eqref{eq:radial_a}. They are
		%
		\label{eq:psiB_solutions}
		\begin{align}
			\psi_B^{(k,+)} &= \frac{\alpha}{2\tilde{k}} r^{l+1} e^{-r^2/4\ell_B^2} \, {}_1F_1\left( 1 - \frac{\ell_B^2 k_{\text{eff}}^2}{2}; l+2; \frac{r^2}{2\ell_B^2} \right), \label{eq:psiB_plus}\\
			\psi^{(k,-)}_{B}&=\frac{\alpha}{2} \tilde{k} r^{-l+1} 
			e^{-r^2{4\ell_B^2}}{}_1F_1\left(1-l-\frac{\ell_{B}^{2}k_{\text{eff}}^{2}}{2};2-l;\frac{r^{2}}{2\ell_{B}^{2}}\right),
			\label{eq:psiB_minus}
		\end{align}
		%
		where $\tilde{k} = k + \alpha m_{\text{eff}} v_F / \hbar$ incorporates the mass correction. 

		\section{Scattering Formalism}
		\label{scattform}
		
		\subsection{Formalism}
		Having established the interior solutions valid within the magnetically active region, we now formulate the scattering problem for an electron incident upon the SQD.
		The incident electron, assumed to propagate in the $+x$-direction with a wavevector $q$ and an energy $E = \hbar v_F q$, is modeled using a plane wave spinor of the form
		\begin{equation}
			\Psi_q^{(i)}(r,\varphi) = \frac{1}{\sqrt{2}} e^{iqr\cos\varphi} \begin{pmatrix} 1 \\ 1 \end{pmatrix}.
			\label{eq:incident}
		\end{equation}
		Employing the Jacobi-Anger expansion, $e^{iz\cos\varphi} = \sum_{l=-\infty}^{\infty} i^l J_l(z) e^{il\varphi}$, this decomposes into angular momentum eigenstates
		\begin{equation}
			\Psi_q^{(i)}(r,\varphi) = \frac{1}{\sqrt{2}} \sum_{l=-\infty}^{\infty} i^l \begin{pmatrix} J_l(qr) e^{il\varphi} \\ i J_{l+1}(qr) e^{i(l+1)\varphi} \end{pmatrix},
			\label{eq:incident_expanded}
		\end{equation}
		where $J_l(x)$ denotes the Bessel function of the first kind.
		In the exterior region where $r > R$, the scattered electron is required to fulfill outgoing-wave boundary conditions. Consequently, the corresponding reflected spinor is \cite{Hewageegana2008}
		\begin{equation}
			\Psi_q^{(r)}(r,\varphi) = \frac{1}{\sqrt{2}} \sum_{l=-\infty}^{\infty} c_l^{(r)} i^l \begin{pmatrix} H_l^{(1)}(qr) e^{il\varphi} \\ i H_{l+1}^{(1)}(qr) e^{i(l+1)\varphi} \end{pmatrix},
			\label{eq:reflected}
		\end{equation}
		where $H_l^{(1)}(x)$ is the Hankel function of the first kind, which for the limit $x \rightarrow \infty$, it takes  the asymptotic form 
		\begin{align}
			H_l^{(1)}(x) \sim \sqrt{2/(\pi x)} e^{i(x - l\pi/2 - \pi/4)}.
		\end{align}
		For the interior region ($r < R$), the electron wave function relies on a superposition of the previously calculated inner solutions, such as
		\begin{widetext}
			\begin{equation}
				\Psi_k^{(t)}(r,\varphi) = \sum_{l=-\infty}^{-1} c_l^{(t)} \begin{pmatrix} \psi_A^{(k,-)}(r) e^{il\varphi} \\ i\psi_B^{(k,-)}(r) e^{i(l+1)\varphi} \end{pmatrix} + \sum_{l=0}^{\infty} c_l^{(t)} \begin{pmatrix} \psi_A^{(k,+)}(r) e^{il\varphi} \\ i\psi_B^{(k,+)}(r) e^{i(l+1)\varphi} \end{pmatrix},
				\label{eq:transmitted}
			\end{equation}
		\end{widetext}
		where $k$ is the wavevector inside the dot, determined by energy conservation.
		
		To ensure physical consistency, the spinor wavefunction must remain continuous across the boundary of the quantum dot at $r = R$, which yields the following matching conditions
		\begin{equation}
			\Psi_q^{(i)}(R,\varphi) + \Psi_q^{(r)}(R,\varphi) = \Psi_k^{(t)}(R,\varphi).
			\label{eq:boundary}
		\end{equation}
		Exploiting orthogonality of the angular factors and conservation of orbital angular momentum in each scattering channel, we obtain the scattering coefficients
		\begin{align}
			c_l^{(t)} &= \frac{i\sqrt{2} \, e^{il\pi/2}}{\pi k R \left[ H_l^{(1)}(qR) \psi_B^{(k,\pm)}(R) - H_{l+1}^{(1)}(qR) \psi_A^{(k,\pm)}(R) \right]}, \label{eq:ct} \\
			c_l^{(r)} &= \frac{-J_l(qR) \psi_B^{(k,\pm)}(R) + J_{l+1}(qR) \psi_A^{(k,\pm)}(R)}{H_l^{(1)}(qR) \psi_B^{(k,\pm)}(R) - H_{l+1}^{(1)}(qR) \psi_A^{(k,\pm)}(R)}, \label{eq:cr}
		\end{align}
		with the superscript $\pm$ selected according to the sign of the quantum number $l$.
		
		\subsection{Scattering Observables}
		\label{subsec:observables}
		
		The physical quantities characterizing the scattering process are derived from the wave functions constructed above for the case of magnetic control of electron scattering in silicene quantum dots. In particular, we evaluate the spatial probability density associated with the charge carrier’s state, which provides direct insight into the localization properties and the formation of quasi-bound states within the quantum dot. The probability current density is derived from spinor solutions, which allow us to study the movement of carriers and the effects of magnetic fields and spin-orbit coupling on it.  These measurements can be used to determine the effectiveness with which quantum dots scatter electrons and the ability of particles to redirect incoming electrons. Combining these two observable factors enables researchers to analyze all aspects of scattering behavior, demonstrating how different confinement methods and interference effects interact with spin- and valley-dependent transport processes in silicene nanostructures.

		Accordingly, the spatial probability density associated with the state of charge carrier is defined as
		\begin{equation}
			\rho(r,\varphi) = \Psi^\dagger(r,\varphi) \Psi(r,\varphi),
			\label{eq:density}
		\end{equation}
		where $\Psi = \Psi^{(t)}$ inside the dot and $\Psi = \Psi^{(i)} + \Psi^{(r)}$ outside.
		The probability density functions form the basis for analyzing the interference patterns that emerge between the incoming waves and the waves bouncing back from the area surrounding the dot. The dot illustrates how magnetic fields and spin-orbit coupling collaborate to generate novel confinement patterns that influence electronic weight distribution. Studying this quantity helps us understand how magnetic confinement interacts with spin-dependent interactions and the geometric features of systems that control the transport and scattering behavior of silicene quantum dots.

		Similarly, the dynamical transport of the system can be evaluated through the probability current density, which takes the following form for Dirac fermions:
		\begin{equation}
			\mathbf{j}(r,\varphi) = v_F \Psi^\dagger(r,\varphi) \bm{\sigma} \Psi(r,\varphi).
			\label{eq:current}
		\end{equation}
		This quantity describes the movement of charge carriers through the system by providing the direction and intensity of the flow.
		Current density shows how electrons move through various parts of the quantum dot by displaying their movement patterns. The area outside the dot shows how incoming waves interact with reflected waves to create specific scattering patterns that move in particular directions. The dot exhibits intricate movement patterns, including multiple current paths and vortex patterns resulting from magnetic field and spin-orbit coupling interactions.
		Analyzing the function $\mathbf{j}(r, \varphi)$ provides a clear visual representation of how electrons move, scatter, and become trapped inside SQDs.

		By examining the far-field radial component of the reflected current, we can extract the differential scattering cross section, which provides detailed information about how electrons are redirected by the quantum dot. Integrating this quantity over all scattering angles allows us to define the scattering efficiency, normalized by the geometric cross section of the quantum dot ($2R$), leading to the expression \cite{Heinisch2013, Schulz2015}
		\begin{equation}
			Q = \frac{4}{qR} \sum_{l=-\infty}^{\infty} |c_l^{(r)}|^2,
			\label{eq:efficiency}
		\end{equation}
		where $q$ is the wave number of the incident electron and $c_l^{(r)}$ are the reflection coefficients of the partial waves.
		Notably, peaks in $Q$ as a function of system parameters indicate the presence of scattering resonances, which correspond to quasi-bound states within the quantum dot. These resonances reflect enhanced confinement and interference effects, providing a direct link between the scattering efficiency and the underlying quantum dynamics of electrons under the influence of magnetic fields and spin–orbit interactions.

		\section{Numerical Analysis}
		\label{sec:results}

		After analytically determining the key physical quantities of the system, we proceed with a numerical investigation of electron scattering on a silicene quantum dot under a constant perpendicular magnetic field. 
		We focus our analysis on the spatial distributions of the probability density  $\rho(r,\varphi)$, the current density $\mathbf{j}(r,\varphi)$, and the scattering efficiency $Q$ in the vicinity of the quantum dot. We pay special attention to the role of intrinsic spin-orbit coupling in silicene, as it modifies the confinement landscape and introduces spin- and valley-dependent features in transport.
		By systematically varying the magnetic field and energy parameters, we uncover how quasi-bound states, interference patterns, and circulating current vortices emerge, providing insight into the interplay between magnetic confinement, SOC, and quantum interference in shaping the scattering and transport behavior of the system.
		
		\begin{figure}[ht!]
			\centering
			\includegraphics[scale=0.55]{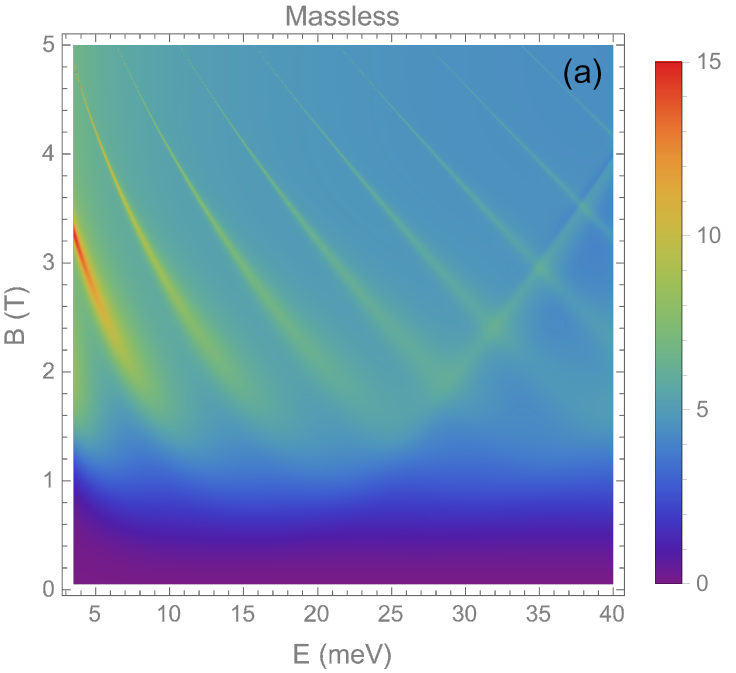}\\
			\includegraphics[scale=0.55]{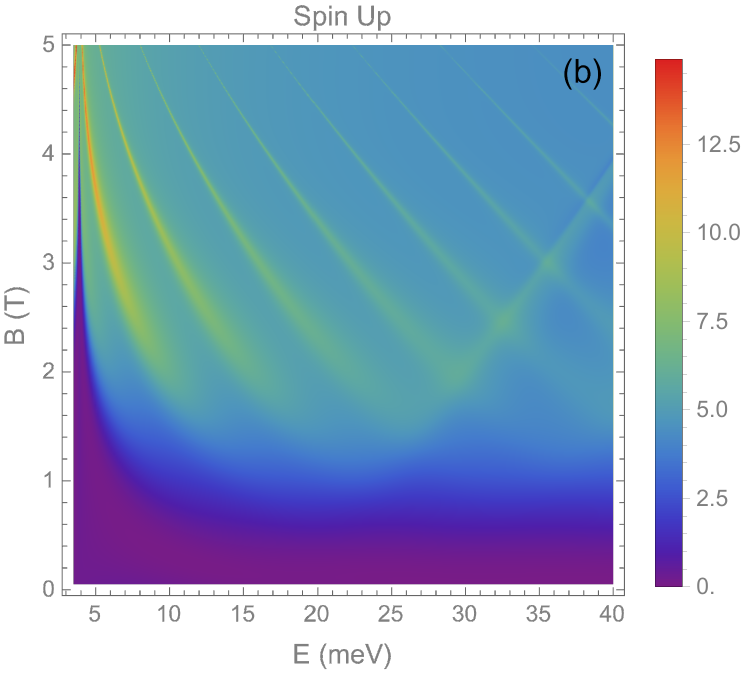}\\
			\includegraphics[scale=0.55]{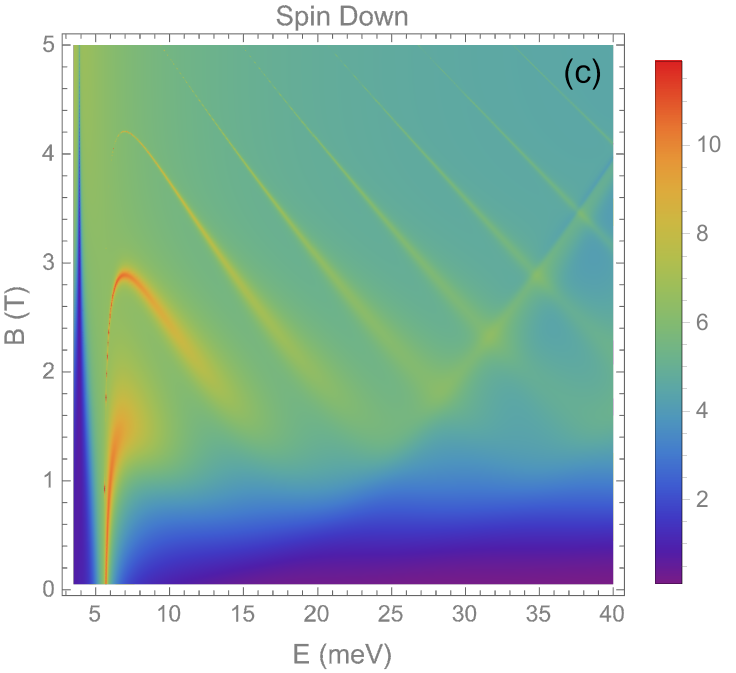}
			\caption{Scattering efficiency $Q$ versus the incident energy $E$ and magnetic field $B$ for a circular SQD of radius $R=50\,\mathrm{nm}$, three spin–orbit coupling intensities, and spin projections (a):  $\lambda_{\mathrm{SO}}=0$, (b): $\lambda_{\mathrm{SO}}=3.9\,\mathrm{meV}$ and $s_z=+1$, and (c): $\lambda_{\mathrm{SO}}=3.9\,\mathrm{meV}$ and $s_z=-1$.} \label{fig1}
		\end{figure}
		
		Figure~\ref{fig1} shows the scattering efficiency $Q$ as a function of the magnetic field $B$ and incident energy $E$ for three different SOC strengths and spin projections, demonstrating a fundamental difference between the behavior of graphene and silicene. In Fig.~\ref{fig1}a, corresponding to the massless limit case ($\lambda_{SO} = 0$) characteristic of pure graphene, magnetic resonances appear as distinct lobes that shift toward higher energies as the magnetic field increases. This behavior reflects the classical orbital confinement limited by the Klein tunneling effect, in agreement with recent observations on magnetic quasi-bound states in graphene quantum dots \cite{Pena2022}.  The introduction of the intrinsic strong SOC in silicene ($\lambda_{SO} = 3.9$ meV), shown in Figs.~\ref{fig1}(b,c), dramatically transforms this topology by first opening a clear energy gap at low energies. This confirms the role of SOC as an effective potential barrier that suppresses transmission, an effect analogous to that of the previously studied artificial mass term \cite{Elazarmasse}. 
		Furthermore, the resonant structures observed in Figs.~\ref{fig1}(b,c) are considerably more intense and sharper than in Fig.~\ref{fig1}a. This enhancement shows that introducing an effective mass eliminates leaks due to the Klein tunneling, allowing carrier localization to be achieved that is far superior to that documented for magnetic flux confinement alone \cite{ElAzarflux}.
		Finally, comparison between the spin-up (Fig.~\ref{fig1}b) and spin-down (Fig.~\ref{fig1}c) configurations reveals a clear asymmetry in the resonance distribution. This lifting of degeneracy, resulting from interference between the spin-dependent mass term and the magnetic vector potential, demonstrates the intrinsic ability of the SQD to filter spins. Such property is usually only obtained in graphene under the action of complex polarized laser fields \cite{Bouhlallaser}.

		\begin{figure}[ht]
			\centering
			\includegraphics[scale=0.55]{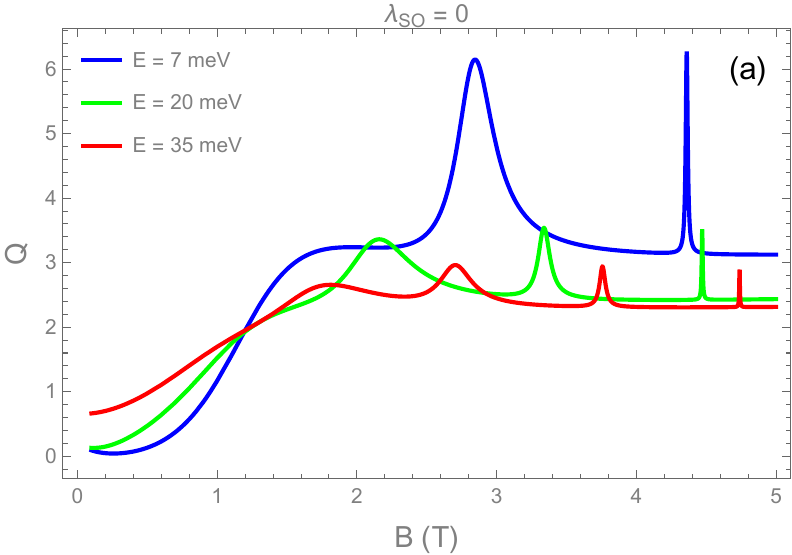}
			\includegraphics[scale=0.55]{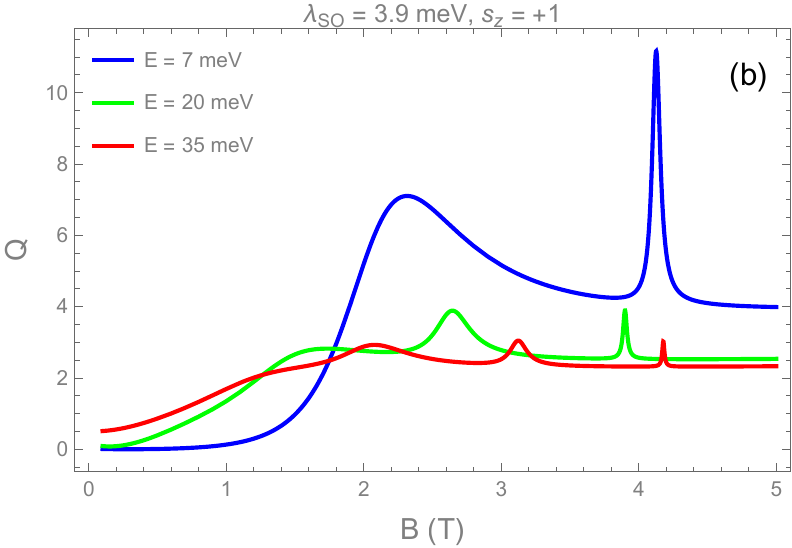}
			\includegraphics[scale=0.55]{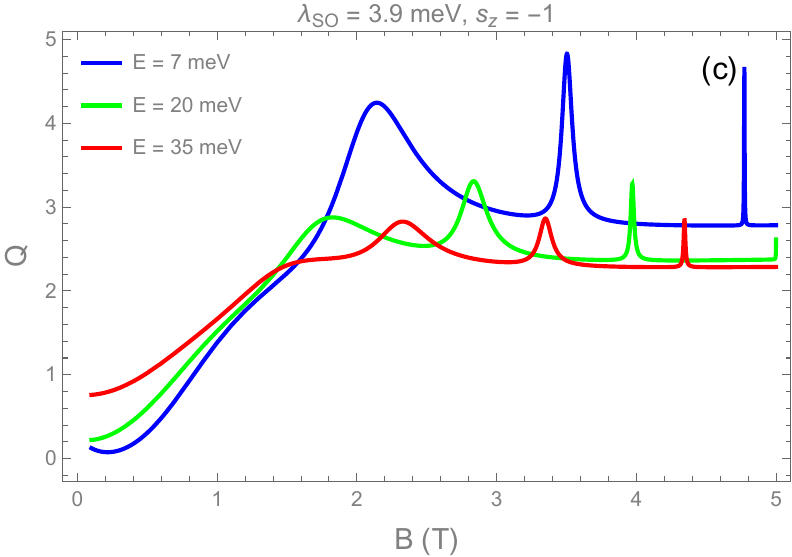}
			\caption{Scattering efficiency $Q$ as a function of the magnetic field $B$ for a circular SQD of radius $R=50\,\mathrm{nm}$,  three incident energies, spin–orbit coupling intensities, and spin projections. 
			} \label{fig2}
		\end{figure}
		
		To measure the strength of magnetic confinement in SQDs with constant incident energy, Figure~\ref{fig2} tracks the scattering efficiency $Q$ as a function of the magnetic field $B$ on a quantum dot with a fixed radius $R=50$ nm and for three incident energies $E$. Without mass (Fig.~\ref{fig2}a), which reproduces the behavior of pure graphene, the increase in the magnetic field causes $Q$ to oscillate. These variations reflect the incipient formation of quasi-bound states. However, the asymmetric and broadened shape of these structures reveals that certain scattering modes, notably the $l=2$ mode, are excited in a non-resonant manner. This partial excitation, combined with the inevitable leakage due to the Klein tunneling effect, shows that magnetic trapping remains fundamentally weak for massless fermions \cite{Pena2022}. 
		In contrast, Figs.~\ref{fig2}(b,c) illustrate how the inclusion of spin–orbit coupling in silicene drastically changes the scattering landscape.
		The SOC ($\lambda_{\mathrm{SO}}=3.9$ meV) induces an effective mass that blocks transmission at normal incidence. Instead of simple fluctuations, we now observe the emergence sharp, well-isolated resonance peaks. This perfect spectral structure indicates that the scattering modes, specifically $l=3$ and $l=4$, are now excited in a purely resonant manner.
		signaling strong electron trapping within the SQD. This results in a substantial increase in the electron lifetime inside the quantum dot, far surpassing what can be achieved through purely magnetic confinement\cite{Pena2022,Elazarmasse,ElAzarflux}.

		\begin{figure}[ht]
			\centering
			\includegraphics[scale=0.55]{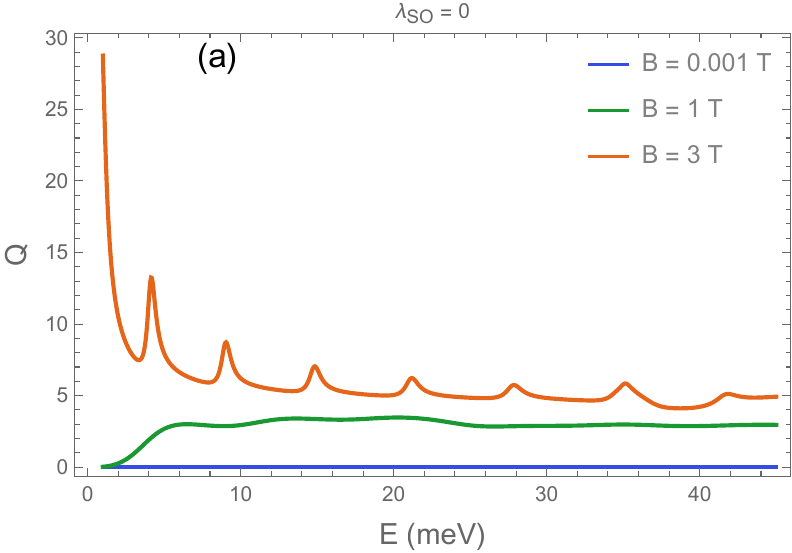}
			\includegraphics[scale=0.55]{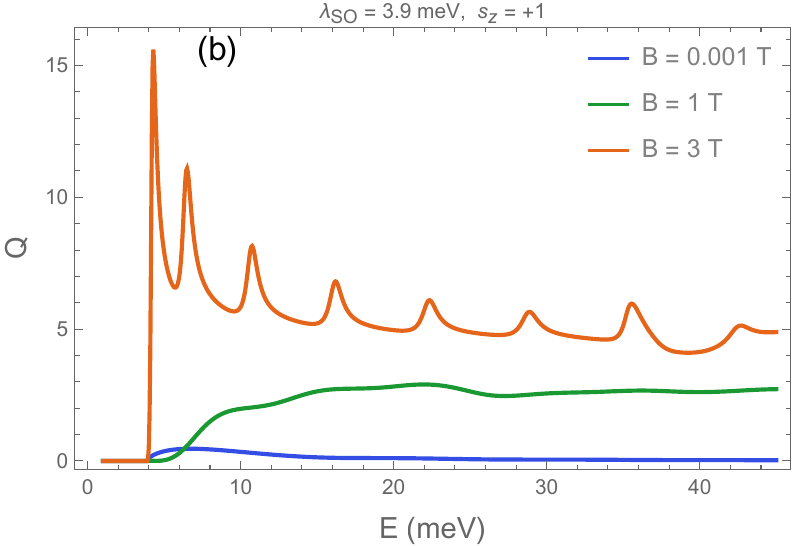}
			\includegraphics[scale=0.55]{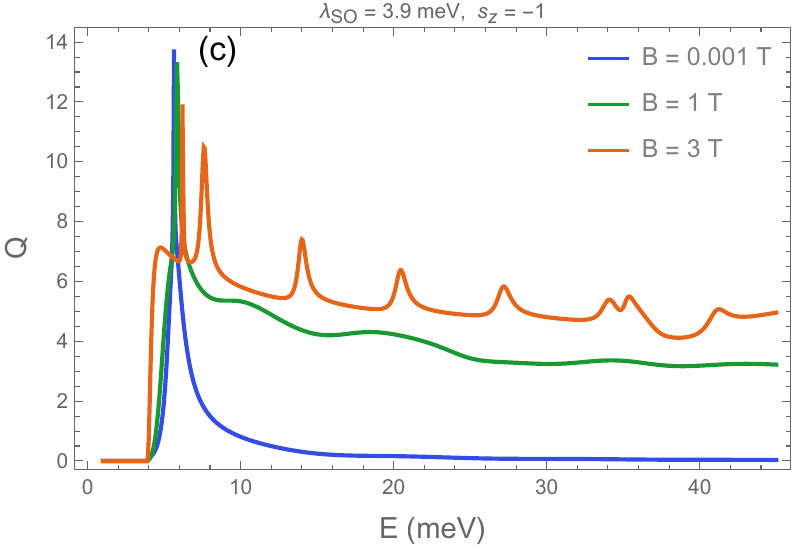}
			\caption{Scattering efficiency $Q$ 
				as a function of the incident energy $E$ for a circular SQD of radius $R=50$ nm,
				three magnetic fields $B$,  spin–orbit coupling intensities, and spin projections. 
			} \label{fig3}
		\end{figure}

		Figure~\ref{fig3} shows the spectral evolution of the scattering efficiency $Q$  as a function of the incident energy $E$  for three distinct magnetic field intensities. The spectrum immediately exhibits sharp, well-defined peaks, which are unmistakable signatures of the formation of quasi-bound states within the structure. A comparison between the massless case in Fig. \ref{fig3}a and the silicene configurations in Figs.~\ref{fig3}(b,c) reveals a fundamental difference. In the graphene limit ($\lambda_{SO}=0$), the resonances appear relatively broad and weak, reflecting significant electron leakage via the Klein tunneling effect. In contrast, the activation of SOC in silicene generates peaks much greater sharpness and intensity. This spectral sharpness shows that the gap opened by SOC acts as an effective barrier, considerably extending the lifetime of trapped carriers, in a way similar to mechanisms induced by polarized laser fields \cite{Bouhlallaser}.
		The influence of the magnetic field is also clearly visible,
		increasing $B$ from 0.001 T to 3 T leads to a systematic shift of the resonances towards higher energies. The blue shift confirms that magnetic confinement, characterized by the magnetic length $l_B$, overlaps with geometric confinement to further compress the electron wave function. Finally, a comparison between the spin-up (Fig.~\ref{fig3}b) and spin-down (Fig.~\ref{fig3}c) spectra reveals 
		noticeable differences in both amplitude and energy position
		for the same magnetic field. This spin-specific dependence confirms that the interaction between the external magnetic field and the intrinsic SOC lifts the degeneracy of the states, enabling selective control of spin polarization through tuning of the magnetic field \cite{ElAzarflux}.
		
		\begin{figure}[ht]
			\centering
			\includegraphics[scale=0.55]{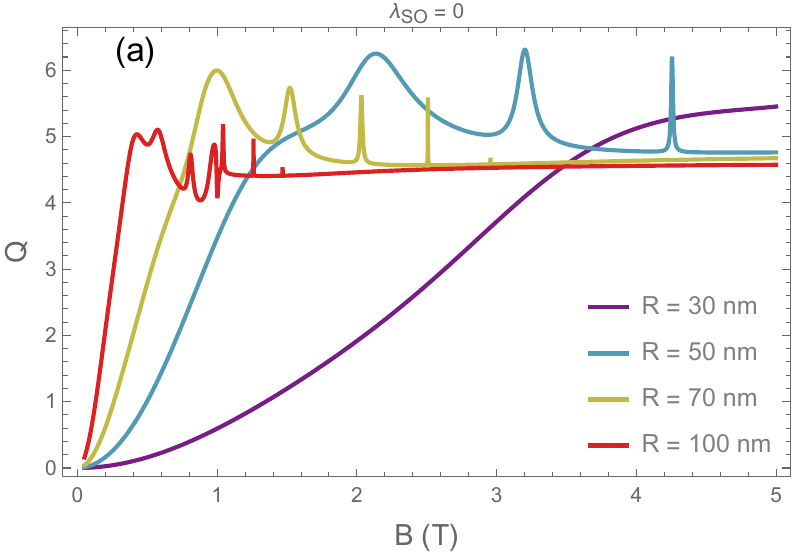}
			\includegraphics[scale=0.55]{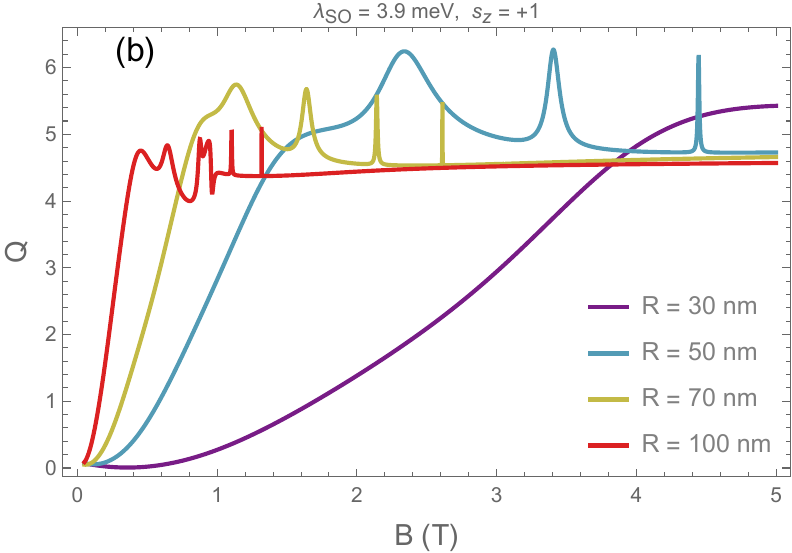}
			\includegraphics[scale=0.55]{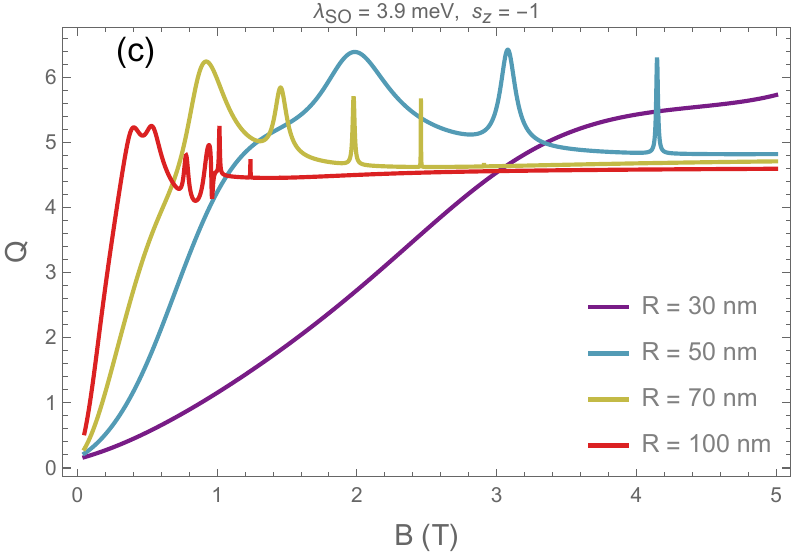}
			\caption{Scattering efficiency $Q$ as a function of the magnetic field $B$ for $E=20$ meV, and four SQD sizes $R = (30, 50, 70, 100)$ nm. (a): Massless, (b): spin-up, and  (c): spin-down.
			} \label{fig4}
		\end{figure}

		Figure~\ref{fig4} shows the dependence of scattering efficiency $Q$ on magnetic field $B$ for different quantum dot dimensions, at a fixed incident energy ($E=20$ meV). It highlights the interplay between two characteristic length scales: the magnetic length $l_B$ and the geometric radius of the system. In Fig. \ref{fig4}a (massless limit), the expected oscillations of the quasi-bound magnetic states specific to graphene are observed. However, these structures remain broad and weakly defined for low radii, demonstrating the intrinsic limitations of massless carrier confinement in small geometries, where the Klein tunneling effect promotes electron escape \cite{Pena2022}. The activation of SOC significantly modifies this behavior, as illustrated in Figs. \ref{fig4}(b,c). The opening of the gap stabilizes the resonances, sharper, more pronounced, and more numerous, particularly for large sizes ($R=100$ nm). This behavior is indicative of the enhanced capacity of the magnetic well to accommodate additional orbital modes as the radius increases, for a given field. Furthermore, a clear spin-dependent separation
		emerges: for an identical radius and magnetic field (e.g., $R=70$ nm), the resonance peaks differ in both position and amplitude depending on the spin orientation. This distinction is of particular importance because it demonstrates that the magnetic field fulfills a dual role, spatial confinement and spin selection, thereby acting as a lever for polarization control, a duality that is impossible to achieve in intrinsic graphene without resorting to elaborate doping or laser irradiation strategies \cite{Bouhlallaser}.

		\begin{figure*}[htbp]
			\centering
			\includegraphics[scale=0.44]{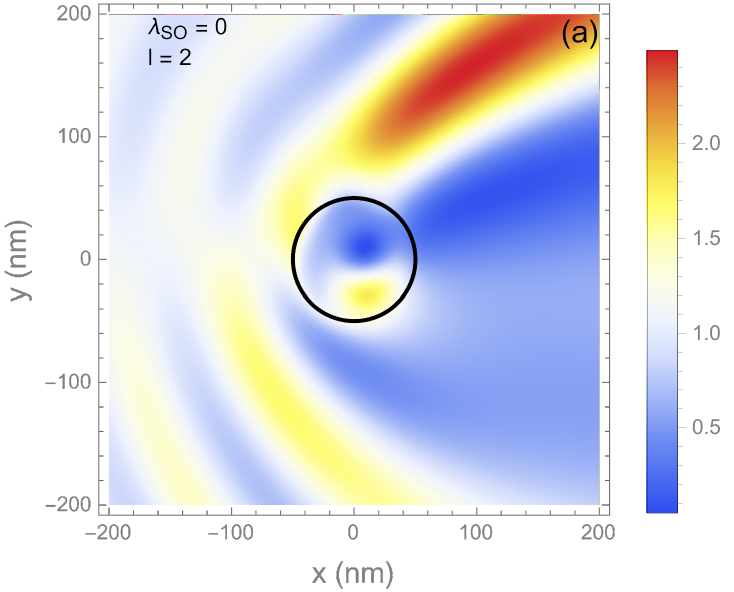}\includegraphics[scale=0.44]{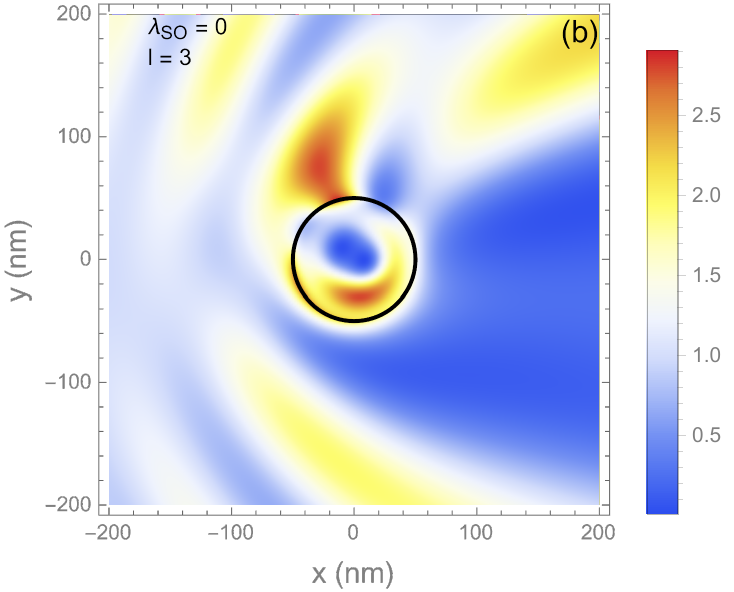}\includegraphics[scale=0.438]{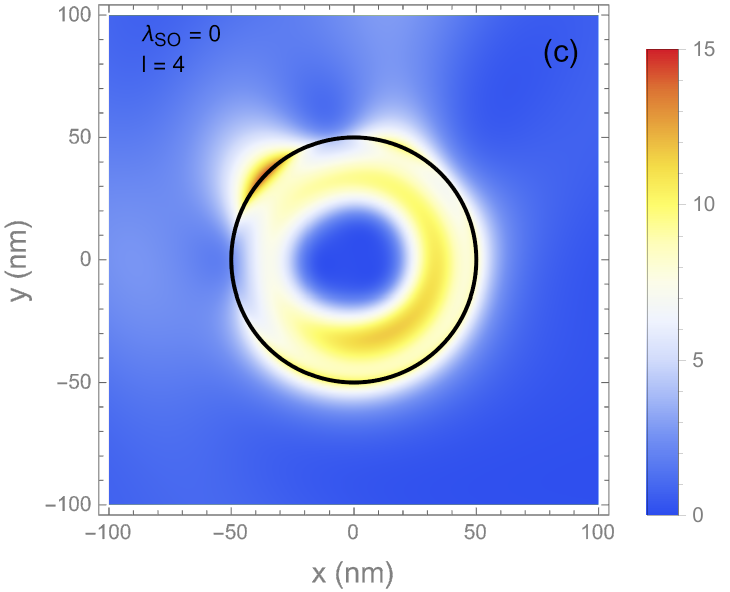}\\
			\includegraphics[scale=0.44]{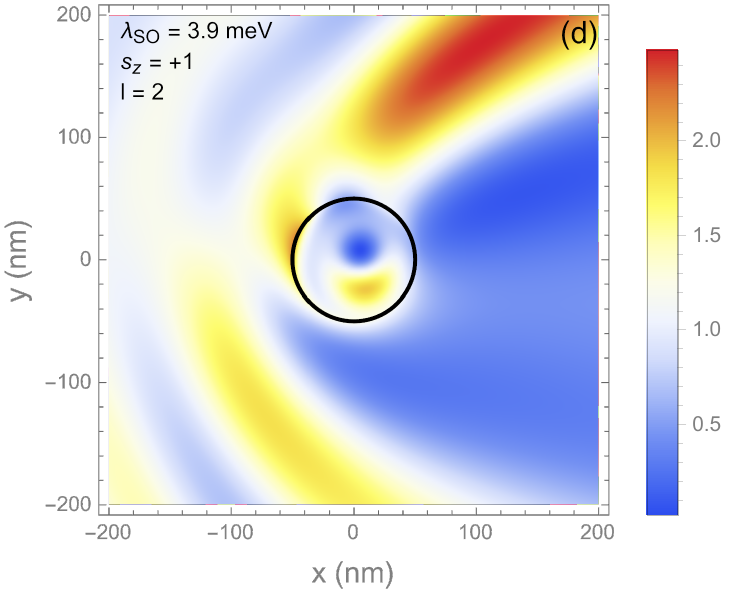}\includegraphics[scale=0.44]{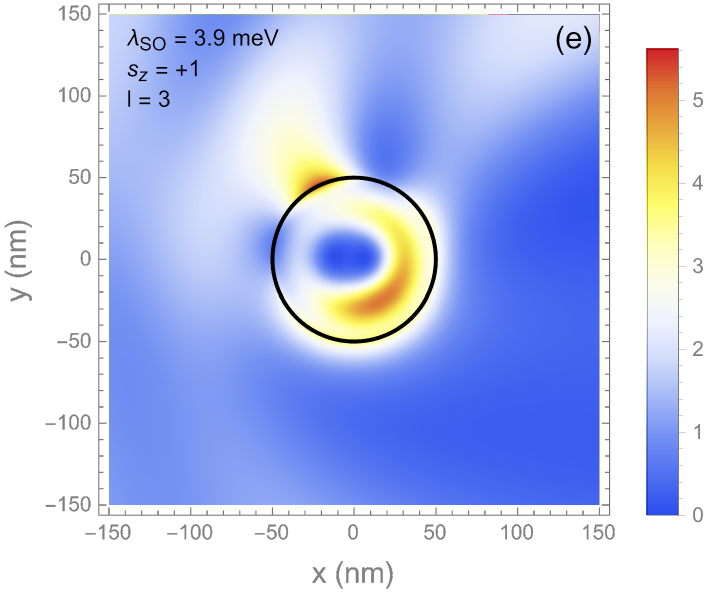}\includegraphics[scale=0.44]{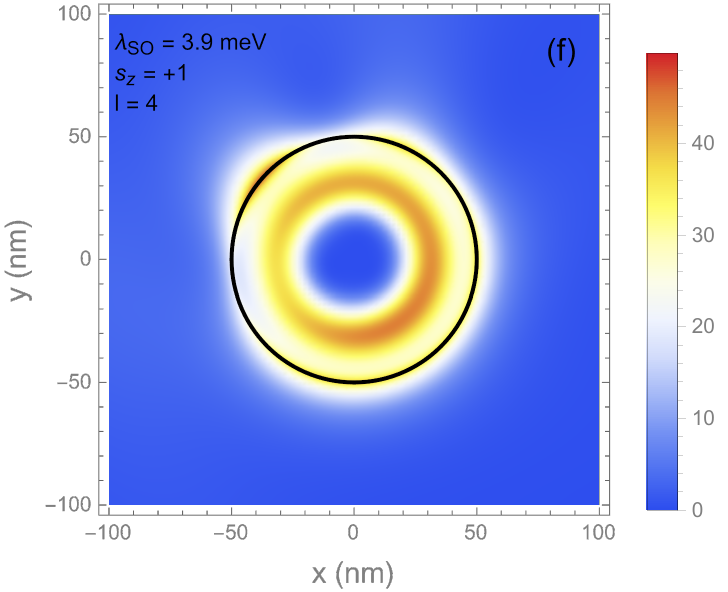}\\
			\includegraphics[scale=0.44]{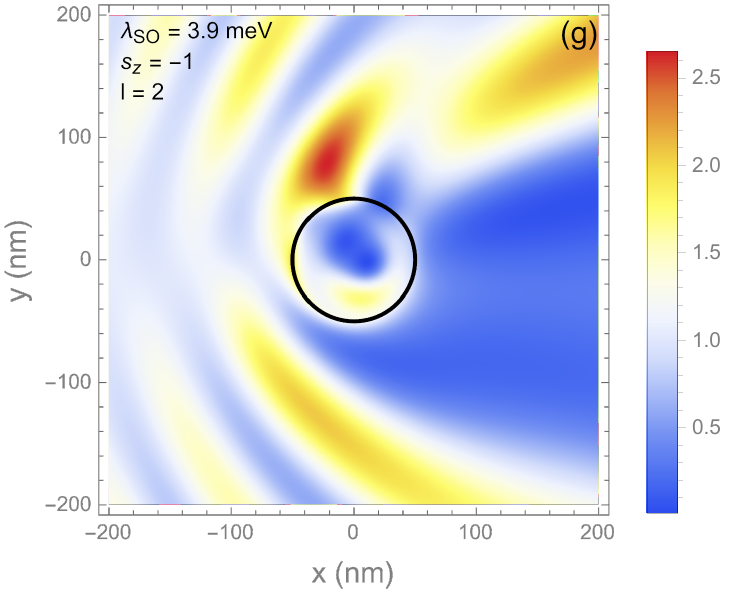}\includegraphics[scale=0.44]{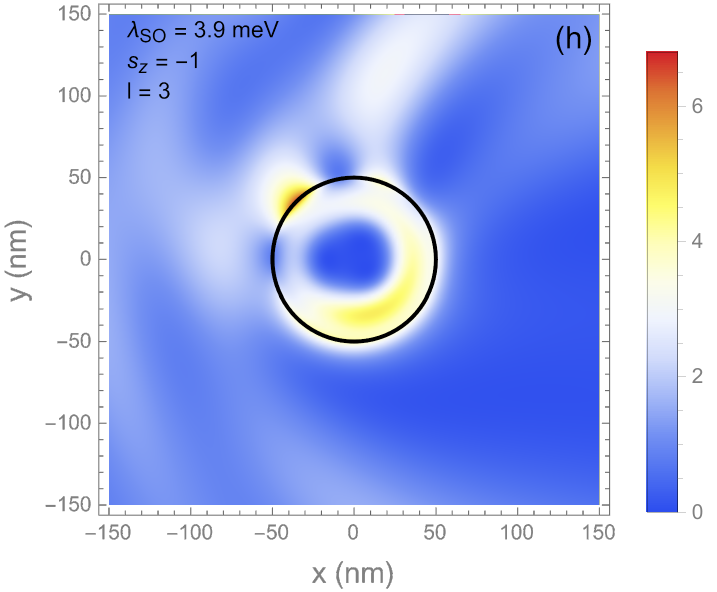}\includegraphics[scale=0.44]{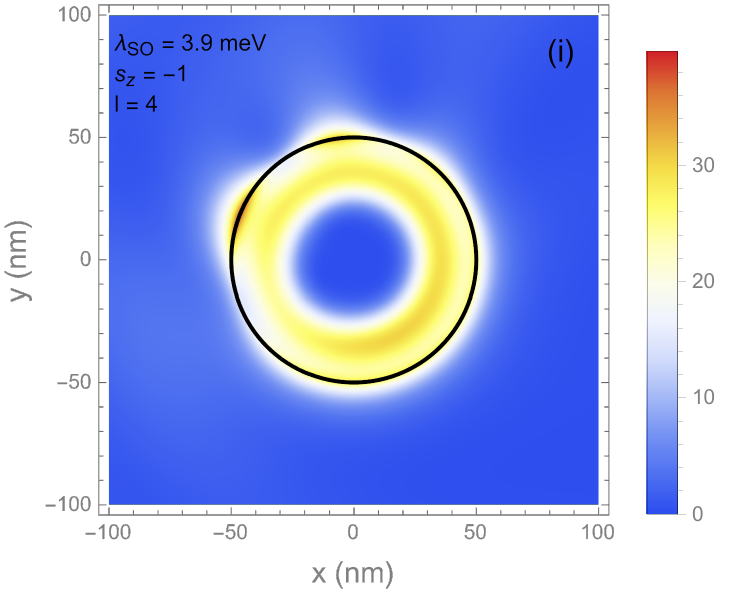}
			\caption{
				Real-space representation of the probability density $\rho(r,\varphi)$ for electron scattering by a magnetically driven SQD at an incident energy $E = 20$~meV. The three columns correspond to the angular momentum quantum numbers $l = 2$, $3$, and $4$, respectively, while the rows represent the massless case, the spin-up state, and the spin-down state.
				The magnetic-field values for each panel are
				(a): $B = 1.18$~T, (b): $B = 2.17$~T, (c): $B = 3.34$~T,
				(d): $B = 1.60$~T, (e): $B = 2.65$~T, (f): $B = 3.90$~T,
				(g): $B = 1.83$~T, (h): $B = 2.83$~T, and (i): $B = 3.97$~T.
				The boundary of the SQD is indicated by the black circle.
			} \label{fig6}
	\end{figure*}

\begin{figure*}[ht!]
		\centering
		\includegraphics[scale=0.435]{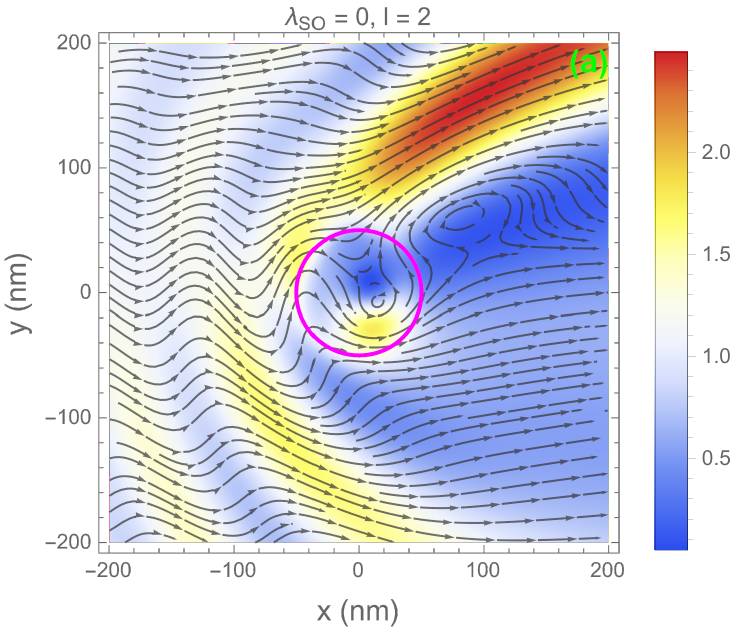}\includegraphics[scale=0.435]{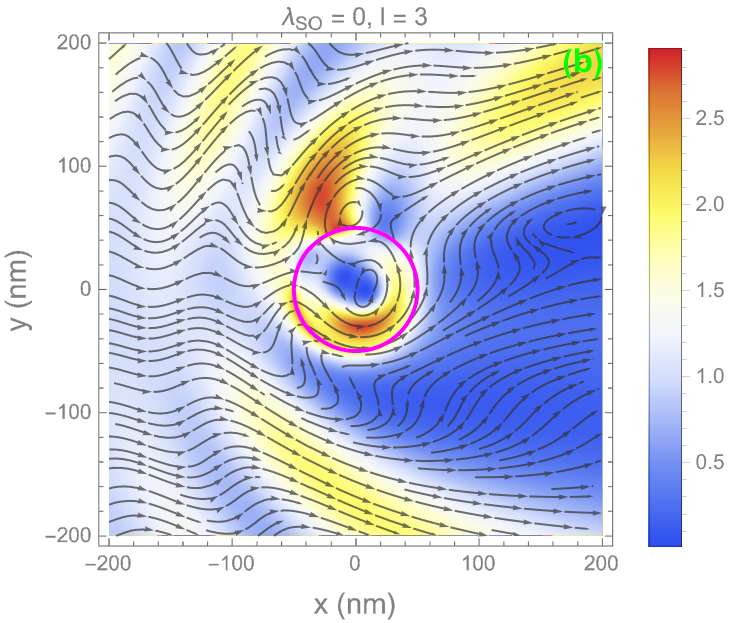}\includegraphics[scale=0.435]{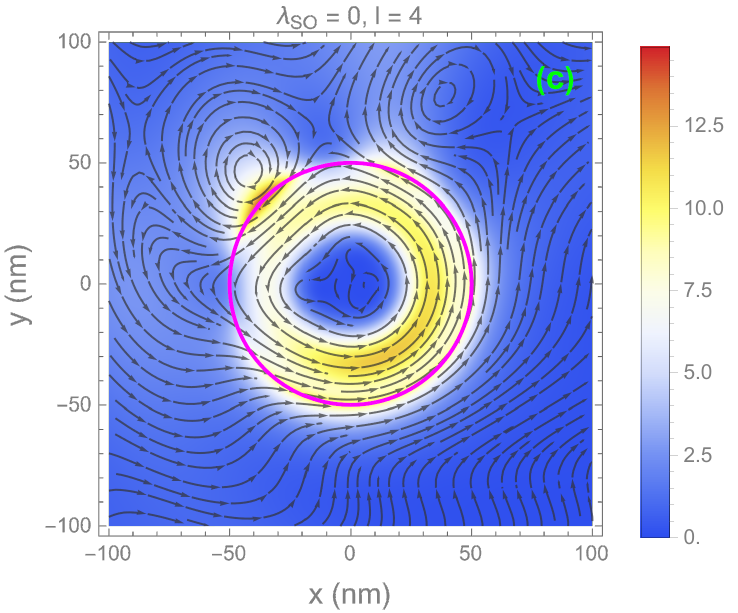}\\
		\includegraphics[scale=0.44]{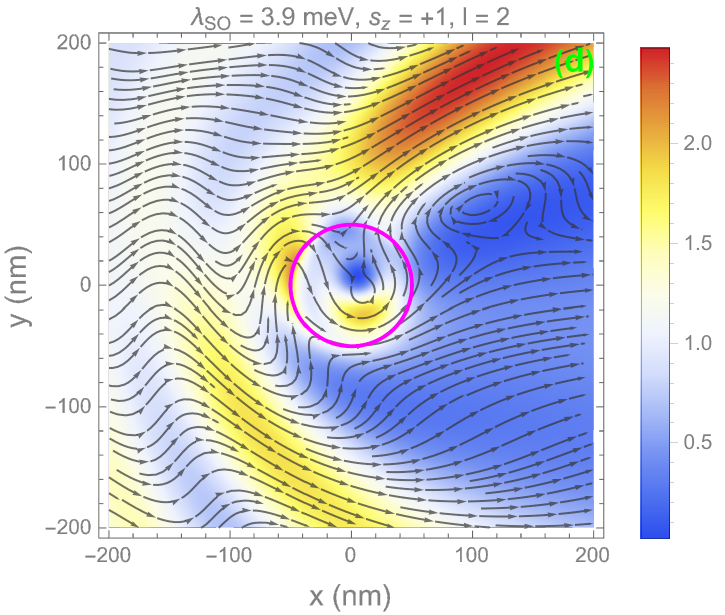}\includegraphics[scale=0.44]{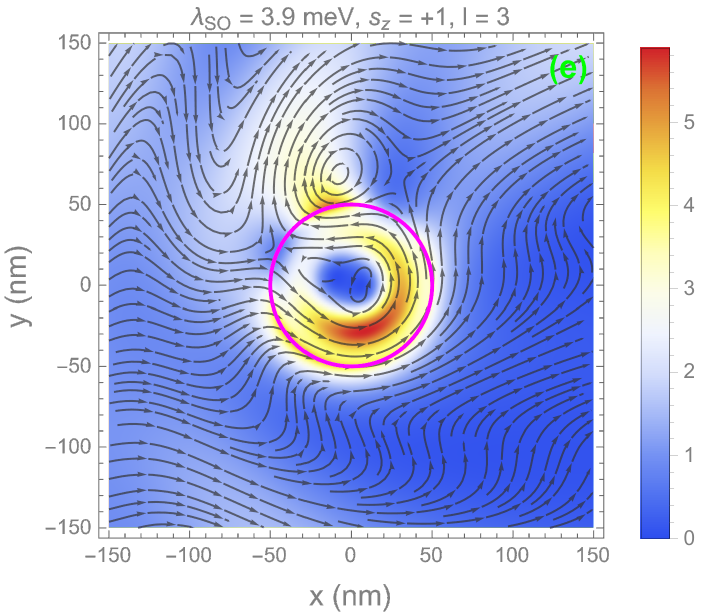}\includegraphics[scale=0.44]{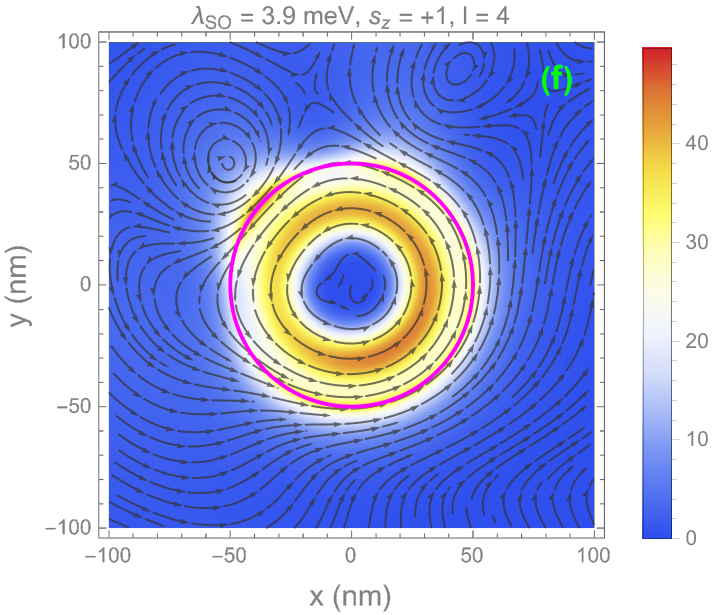}\\ \includegraphics[scale=0.44]{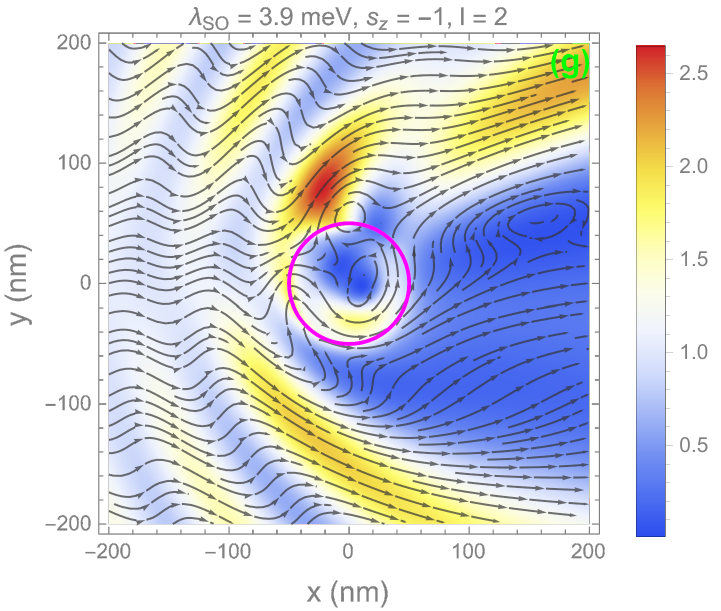}\includegraphics[scale=0.44]{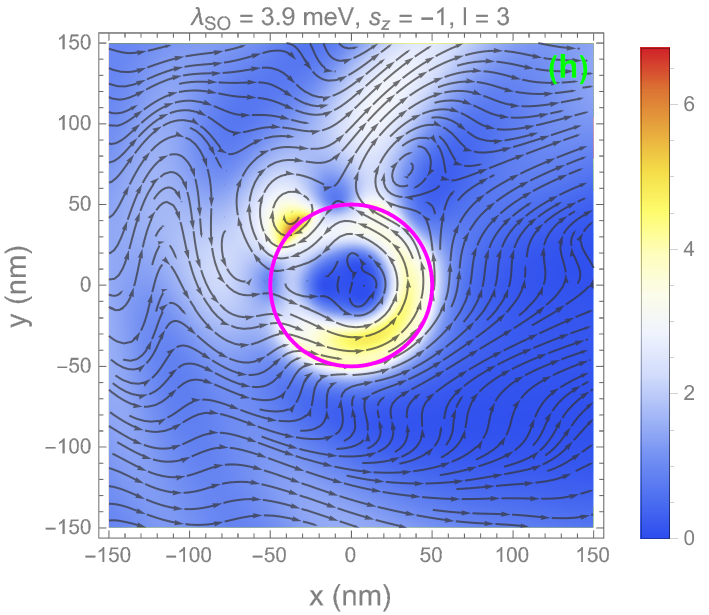}\includegraphics[scale=0.44]{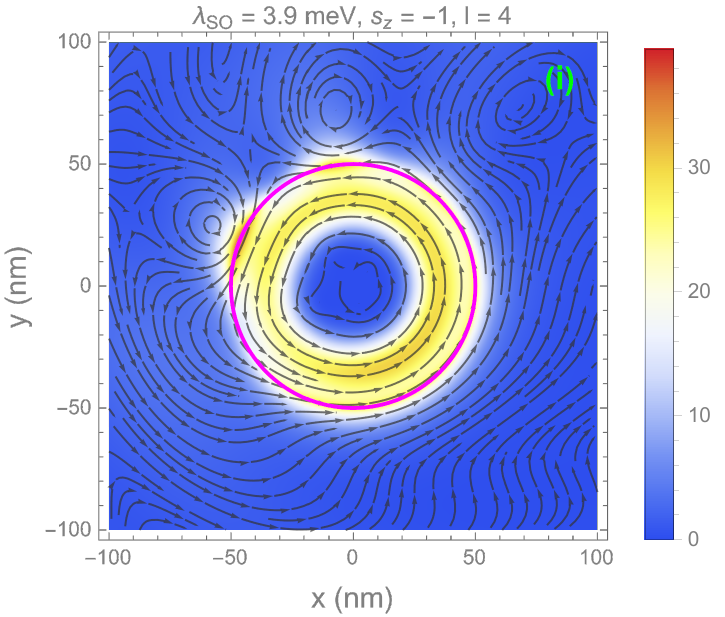}
		\caption{Real-space representation of the current density $\mathbf{j}(r,\varphi)$ for electron scattering by a magnetically driven SQD at an incident energy $E = 20$~meV. The three columns correspond to the angular momentum quantum numbers $l = 2$, $3$, and $4$, respectively, while the three rows represent the massless case, the spin-up state ($s_z = +1$), and the spin-down state ($s_z = -1$). The magnetic-field values for each panel are: (a) $B = 1.18$~T, (b) $B = 2.17$~T, (c) $B = 3.34$~T, (d) $B = 1.60$~T, (e) $B = 2.65$~T, (f) $B = 3.90$~T, (g) $B = 1.83$~T, (h) $B = 2.83$~T, and (i) $B = 3.97$~T. The boundary of the SQD is indicated by the purple circle.}
	\label{fig7}

	\end{figure*}

		The spatial distributions of probability density and current density have been presented separately to help understand how the scattering process occurs. Figure~\ref{fig6} shows only the probability density \(\rho(r,\varphi)\) which enables to see how the quasi-bound states inside the SQD exhibit different localization patterns and interference structures and symmetrical properties. The spatial distributions in Fig~\ref{fig7} are combined with the current density vector field \(\mathbf{j}(r,\varphi)\) which allows for the study of flow dynamics that develop circulating currents and vortex structures that magnetic fields and spin-orbit coupling create. The two figures present different information which helps to understand the confinement properties and transport mechanisms through their step-by-step analysis.

	In order to directly visualize the formation and stability of the quasi-bound states discussed in Figure~\ref{fig2}, Figure~\ref{fig6} maps the electronic probability density in real space, both inside and around the SQD (with the boundary marked by the black circle), for a fixed incident energy of $E=20$ meV. The columns trace the evolution of this distribution for different values of the angular momentum ($l=2, 3, 4$), while the rows highlight the respective influence of the effective mass and spin orientation.
	Analysis of the first row, Figs.~\ref{fig6}(a,b,c), corresponding to the massless limit (graphene-like case), shows that although the magnetic field induces some charge accumulation inside the dot, a significant fraction of the density leaks outside the geometric boundaries. This is clearly visible in Fig.~\ref{fig6}a, where the incident electron forms diffraction-like fringes and largely bypasses the quantum dot. This spatial delocalization is a direct signature of Klein tunneling, which prevents effective confinement of massless Dirac fermions, particularly for low angular momentum modes, in agreement with previous studies \cite{Pena2022}.
	In contrast, the introduction of SOC in silicene (second and third rows) profoundly modifies this behavior. For the spin-up configuration, Figs.~\ref{fig6}(d,e,f),  and spin-down configuration, 
	Figs.~\ref{fig6}(g,h,i), the probability density becomes strongly localized inside the quantum dot, especially for higher angular momentum states ($l=3,4$). These states exhibit well-defined interference patterns with negligible leakage, demonstrating that the SOC-induced mass term acts as an efficient confinement barrier.
	Furthermore, comparison between the rows reveals a clear lifting of spin degeneracy. For a given angular momentum mode, to achieve a similar confinement, different magnetic fields are required for each spin orientation. For example, $B = 1.6$ T is required for spin-up while $B = 1.83$ T  for spin-down in the case of $l = 2$. This spatial and magnetic asymmetry confirms that the SQD can operate as a tunable spin filter controlled by the applied magnetic field.
	Finally, it is worth noting that the confinement of higher angular momentum modes ($l=3,4$) requires stronger magnetic fields, which reduce the cyclotron radius and force the wave function to remain within the geometric boundaries of the quantum dot.
	
	Figure~\ref{fig7} gives a more intuitive picture of how these quasi-bound states actually behave by showing the spatial distribution of the current density—a kind of “map” of how electrons flow—for the same resonances discussed earlier. Let us first consider the massless case in Figs.~\ref{fig7}(a,b,c). Here, the current lines are quite spread out, and electrons cross the boundary of the quantum dot with little resistance. This clearly illustrates the Klein tunneling effect: carriers are not confined and can easily escape the structure \cite{Pena2022}.
	The introduction of SOC in silicene results in a substantial transformation of the current situation according to the findings presented in Figs.~\ref{fig7}(d,e,f,g,h,i). The current does not extend outward but instead creates distinct closed loops, which especially occur in the $l=3$ and $l=4$ modes that receive resonant excitation according to the previous observation in Fig.~\ref{fig2}.
	In this regime, almost no current leaks outside the dot, indicating that the SOC-induced mass effectively suppresses transmission and traps the electrons in stable circular motion.
	Looking across the columns, one can also see that for higher angular momentum states ($l=3$ and $l=4$), these loops move closer to the edge of the quantum dot. This reflects the stronger “centrifugal” tendency of the electrons, meaning that higher magnetic fields are needed to keep their trajectories confined within the structure.
	The spin-up and spin-down cases show their first differences, which appear to be small yet important. The combination of SOC with the magnetic field produces a minor adjustment to both the current loop patterns and their movement direction, which demonstrates that the transportation process depends on spin state. This highlights an important feature of the system: it can be used to control and even filter spin currents in a very precise way.

	\section{Conclusions}\label{sec:conclusions}

	We have theoretically studied the phenomenon of electron scattering through a silicene quantum dot subjected to a perpendicular magnetic field. By solving the low-energy Dirac equation, we have analytically determined the scattering coefficients, which allowed us to evaluate the scattering efficiency, as well as the spatial distributions of the probability and current densities. This combined analytical and numerical approach provides a complete picture of the scattering dynamics and the underlying confinement mechanisms.
	
	In the massless limit, analogous to pure graphene, magnetic confinement alone remains fundamentally limited by the Klein tunneling effect. The non-resonant excitation of certain scattering modes leads to broad and weak resonances, accompanied by significant electron leakage. This behavior highlights the intrinsic difficulty of confining massless Dirac fermions using magnetic barriers only. In contrast, our results clearly demonstrate that the intrinsic spin–orbit (SOC) coupling of silicene profoundly alters this scenario. By opening a finite energy gap, SOC introduces an effective mass that suppresses transmission, thereby enabling true confinement. As a consequence, we observe the emergence of sharp and well-defined resonance peaks, which are clear signatures of purely resonant excitation and long-lived quasi-bound states.
	This strong enhancement of confinement is further confirmed by the real-space analysis. The probability density maps reveal a pronounced localization of the electronic states inside the quantum dot, while the current density distributions show the formation of stable closed vortices. These circulating currents provide a direct visualization of the trapped states and demonstrate the near absence of leakage outside the dot. Together, these results establish that the interplay between magnetic confinement and SOC leads to a highly efficient trapping mechanism, far exceeding what can be achieved in graphene-based systems.

	Our results highlight the central role of the magnetic field in shaping the behavior of the system. It not only controls the confinement strength but also determines the spectral and spatial features of the quasi-bound states. The tunability of the system arises from the interplay between the magnetic length and the geometric confinement, as evidenced by the systematic shift of resonance patterns with increasing magnetic field.
	The system exhibits spin degeneracy reduction through the combination of external magnetic field and intrinsic SOC effects. The resonance positions and intensities differ significantly for spin-up and spin-down electrons, leading to a clear spin-dependent response. The silicene quantum dot exhibits an asymmetric distribution of its properties, which enables it to function as a natural spin filter. The implementation of magnetic field adjustments allows users to precisely control the movement of carriers based on their specific spin direction, which creates exciting new possibilities for developing spintronics and valleytronics devices that can be precisely controlled.

	Silicene has been successfully synthesized on several substrates, including 
	Ag(111), ZrB$_2$, and Ir(111) surfaces 
	\cite{Vogt2012,Fleurence2012,Meng2013}, and its low-buckled structure 
	naturally gives rise to the spin--orbit coupling that is central to our 
	study. A localized perpendicular magnetic field can be generated 
	experimentally using a magnetized scanning tunneling microscope (STM) tip 
	or by patterning ferromagnetic strips on top of the silicene sheet 
	\cite{DellAnna2011,Ramezani2011}. The quantum dot geometry considered 
	here is defined by the spatial profile of this localized field, which 
	determines the dot radius $R$ and field strength $B$. Both parameters 
	are continuously tunable in experiment, making the proposed configuration 
	practically accessible.
	
	The spatial distributions of probability and current densities predicted 
	in this work are directly related to the local density of states (LDOS) 
	and local current maps, both of which are measurable using scanning 
	tunneling spectroscopy (STS) and STM-based current imaging techniques 
	\cite{Crommie1993,Zhao2015}. Quasibound state resonances manifest as 
	sharp peaks in the differential conductance $dI/dV$ as a function of 
	energy, and their spin-selective nature can in principle be probed using 
	spin-polarized STM \cite{Wiesendanger2009}. The scattering efficiency, 
	which we compute as a function of incident energy and magnetic field 
	strength, is accessible through transport measurements in gated silicene 
	devices.
	
	We acknowledge that fabrication imperfections and substrate-induced 
	disorder represent the primary experimental challenges. Silicene is 
	sensitive to its substrate environment, and charge inhomogeneity or 
	rippling can broaden the quasibound state resonances and reduce the 
	spin contrast. Nevertheless, the intrinsic spin--orbit gap of silicene 
	($\lambda_{\rm SO} \approx 3.9$ meV \cite{Liu2011}) acts as a natural 
	stabilizing mass term that partially protects the confined states against 
	weak disorder. Furthermore, the sharp resonance peaks predicted in our 
	scattering efficiency maps remain robust over a wide range of magnetic 
	field strengths and dot radii, suggesting that the key physical features 
	of our results are not critically sensitive to small fabrication 
	tolerances.


\begin{thebibliography}{99}
	\bibitem{Loss1998} D. Loss and D. P. DiVincenzo, Phys. Rev. A 57, 120 (1998).
	
	\bibitem{Hanson2007} R. Hanson, L. P. Kouwenhoven, J. R. Petta, S. Tarucha, and L. M. K. Vandersypen, Rev. Mod. Phys. 79, 1217 (2007).
	
	\bibitem{Zwanenburg2013} F. A. Zwanenburg, A. S. Dzurak, A. Morello, M. Y. Simmons, L. C. L. Hollenberg, G. Klimeck, S. Rogge, S. N. Coppersmith, and M. A. Eriksson, Rev. Mod. Phys. 85, 961 (2013).
	
	\bibitem{Kloeffel2013} C. Kloeffel and D. Loss, Annu. Rev. Condens. Matter Phys. 4, 51 (2013).
	
	\bibitem{Qammar2024} M. Qammar, M. J. H. Tan, P. Ding, J. Ge, Y. Chan, and J. E. Halpert,
	Nano Research 17, 10426 (2024).
	
	\bibitem{CastroNeto2009} A. H. Castro Neto, F. Guinea, N. M. R. Peres, K. S. Novoselov, and A. K. Geim, Rev. Mod. Phys. 81, 109 (2009).
	
	\bibitem{Katsnelson2006} M. I. Katsnelson, K. S. Novoselov, and A. K. Geim, Nat. Phys. 2, 620 (2006).
	
	\bibitem{Stander2009} N. Stander, B. Huard, and D. Goldhaber-Gordon, Phys. Rev. Lett. 102, 026807 (2009).
	
	\bibitem{Banszerus2020} L. Banszerus, A. Rothstein, T. Fabian, S. Möller, E. Icking, S. Trellenkamp, F. Lentz, D. Neumaier, K. Watanabe, T. Taniguchi, F. Libisch, C. Volk, and C. Stampfer, Nano Lett. 20, 7709 (2020). 
	
	\bibitem{Elahi2024} M. M. Elahi, H. Vakili, Y. Zeng, C. R. Dean, and A. W. Ghosh, Phys. Rev. Lett. 132, 146302 (2024).
	
	\bibitem{Downing2011} C. A. Downing, D. A. Stone, and M. E. Portnoi, Phys. Rev. B 84, 155437 (2011).
	\bibitem{Romanovsky2012} I. Romanovsky, C. Yannouleas, and U. Landman, Phys. Rev. B 85, 165434 (2012).
	\bibitem{Matulis2008} A. Matulis and F. M. Peeters, Phys. Rev. B 77, 115423 (2008).
	\bibitem{Bardarson2009} J. H. Bardarson, M. Titov, and P. W. Brouwer, Phys. Rev. Lett. 102, 226803 (2009).
	\bibitem{Cheianov2006} V. V. Cheianov and V. I. Fal'ko, Phys. Rev. B 74, 041403(R) (2006).
	\bibitem{Gutierrez2016} C. Gutiérrez, L. Brown, C. -J. Kim, J. Park, and A. N. Pasupathy, Nat. Phys. 12, 1069 (2016). 
	\bibitem{Novoselov2005} K. S. Novoselov, A. K. Geim, S. V. Morozov, D. Jiang, M. I. Katsnelson, I. V. Grigorieva, S. V. Dubonos, and A. A. Firsov, Nature 438, 197 (2005).
	\bibitem{Zhang2005} Y. Zhang, Y. -W. Tan, H. L. Stormer, and P. Kim, Nature 438, 201 (2005).
	\bibitem{Dean2013} C. R. Dean, L. Wang, P. Maher, C. Forsythe, F. Ghahari, Y. Gao, J. Katoch, M. Ishigami, P. Moon, M. Koshino, T. Taniguchi, K. Watanabe, K. L. Shepard, J. Hone, and P. Kim, Nature 497, 598 (2013).
	\bibitem{Schelter2012} J. Schelter, P. Recher, and B. Trauzettel, Solid State Commun. 152, 1411 (2012).
	\bibitem{Cahangirov2009} S. Cahangirov, M. Topsakal, E. Aktürk, H. Şahin, and S. Ciraci, Phys. Rev. Lett. 102, 236804 (2009).
	\bibitem{Liuspin2011} C.-C. Liu, W. Feng, and Y. Yao, Phys. Rev. Lett. 107, 076802 (2011).
	\bibitem{Drummond2012} N. D. Drummond, V. Zólyomi, and V. I. Fal'ko, Phys. Rev. B 85, 075423 (2012).
	\bibitem{Szafran2018} B. Szafran and D. Żebrowski, Phys. Rev. B 98, 155305 (2018).
	\bibitem{Chen2024EPL} F. Chen, F. Qi, and G. Jin,
	Europhys. Lett. 148, 16001 (2024).
	\bibitem{Gao2024APL} X. Gao, Z. Deng, C. Ma, L. Li, X. Zhang, X. Li, and Z. Zhou, Appl. Phys. Lett. 124, 082406 (2024).
	
	
	\bibitem{Pena2022}  A. Pena, Physica E 141, 115245 (2022).
	\bibitem{Elazarmasse} M. El Azar, A. Bouhlal, and A. Jellal, Comput. Mater. Sci. 231, 112573 (2024).
	
	
\bibitem{garg2022caustical} N. A. Garg, Eur. Phys. J. B 95, 165 (2022).
	
	
	
	\bibitem{LiuHall2011} C. C. Liu, W. Feng, and Y. Yao, Phys. Rev. Lett. 107, 076802 (2011).
	\bibitem{Liu2011} C. C. Liu, H. Jiang, and Y. Yao, Phys. Rev. B 84, 195430 (2011).
	\bibitem{CahangirovPRL2009} S. Cahangirov, M. Topsakal, E. Aktürk, H. Şahin, and S. Ciraci, Phys. Rev. Lett. 102, 236804 (2009).
	\bibitem{Kara2012} A. Kara, H. Enriquez, A. P. Seitsonen, L. C. Lew Yan Voon, S. Vizzini, B. Aufray, and H. Oughaddou, Surf. Sci. Rep. 67, 1 (2012).
	\bibitem{Recher2007} P. Recher, J. Nilsson, G. Burkard, and B. Trauzettel, Phys. Rev. B 76, 235404 (2007).
	\bibitem{Hewageegana2008}  P. Hewageegana and V. Apalkov, Phys. Rev. B 77, 245426 (2008).
	\bibitem{Heinisch2013} R. Heinisch, F. Bronold, and H. Fehske, Phys. Rev. B 87, 155409 (2013).
	\bibitem{Schulz2015} C. Schulz, R. Heinisch, and H. Fehske, Quantum Matter 4, 346 (2015).
	
	
	
	
	
	\bibitem{ElAzarflux}  M. El Azar, A. Bouhlal, A. D. Alhaidari, and A. Jellal, Physica B 675, 415610 (2024).
	\bibitem{Bouhlallaser} A. Bouhlal, M. El Azar, A. Naciri, E. Feddi, and A. Jellal, Physica E 172, 116273 (2025).
	

		\bibitem{Vogt2012}
		P. Vogt, P. De Padova, C. Quaresima, J. Avila, E. Frantzeskakis, 
		M. C. Asensio, A. Resta, B. Ealet, and G. Le Lay,
		Phys. Rev. Lett. {108}, 155501 (2012).
		
		\bibitem{Fleurence2012}
		A. Fleurence, R. Friedlein, T. Ozaki, H. Kawai, Y. Wang, 
		and Y. Yamada-Takamura,
		Phys. Rev. Lett. {108}, 245501 (2012).
		
		\bibitem{Meng2013}
		L. Meng, Y. Wang, L. Zhang, S. Du, R. Wu, L. Li, Y. Zhang, 
		G. Li, H. Zhou, W. A. Hofer, and H.-J. Gao,
		Nano Lett. {13}, 685 (2013).
		
		\bibitem{DellAnna2011}
		L. Dell'Anna and A. De Martino,
		Phys. Rev. B {83}, 155449 (2011).
		
		\bibitem{Ramezani2011}
		M. Ramezani Masir, P. Vasilopoulos, and F. M. Peeters,
		New J. Phys. {11}, 095009 (2011).
		
		\bibitem{Crommie1993}
		M. F. Crommie, C. P. Lutz, and D. M. Eigler,
		Science {262}, 218 (1993).
		
		\bibitem{Zhao2015}
		Y. Zhao, J. Wyrick, F. D. Natterer, J. F. Rodriguez-Nieva, 
		C. Lewandowski, K. Watanabe, T. Taniguchi, L. S. Levitov, 
		N. B. Zhitenev, and J. A. Stroscio,
		{Science} {348}, 672--675 (2015).
		
		\bibitem{Wiesendanger2009}
		R. Wiesendanger,
		{Rev. Mod. Phys.} {81}, 1495--1550 (2009).
		
		
	
	
\end{thebibliography}
\end{document}